\documentclass[final,1p,times,authoryear]{elsarticle}
\usepackage{MR}
\usepackage{amssymb}
\usepackage{subcaption}

\graphicspath{{fig/}}

\journal{J. Computational Physics}

\newcommand{\vect}[1]{\mathbf{#1}}
\newcommand{\tens}[1]{\mathcal{#1}}
\def\rvec{\mathbf{\hat r}}
\def\thvec{\vec{\hat \theta}}
\def\phivec{\vec{\hat \varphi}}

\renewcommand{\vec}{\mathbf}
\begin{document}

\begin{frontmatter}

\title{An algorithm for computing the 2D structure of fast rotating
stars}

\author[add1,add2]{Michel Rieutord}
\author[add1,add2]{Francisco Espinosa Lara\footnote{Present address:
Space Research Group, University of Alcal\'a, 28871 Alcal\'a de Henares, Spain}}
\author[add1,add2]{Bertrand Putigny}

\address[add1]{Universit\'e de Toulouse; UPS-OMP; IRAP; Toulouse, France}
\address[add2]{CNRS; IRAP; 14, avenue Edouard Belin, F-31400 Toulouse,
France}

\begin{abstract}
Stars may be understood as self-gravitating masses of a compressible fluid whose
radiative cooling is compensated by nuclear reactions or gravitational
contraction. The understanding of their time evolution requires the use of
detailed models that account for a complex microphysics including that
of opacities, equation of state and nuclear reactions.  The present
stellar models are essentially one-dimensional, namely spherically
symmetric. However, the interpretation of recent data like the surface
abundances of elements or the distribution of internal rotation have
reached the limits of validity of one-dimensional models because of
their very simplified representation of large-scale fluid flows. In this
article, we describe the ESTER code, which is the first code able to
compute in a consistent way a two-dimensional model of a fast rotating
star including its large-scale flows. Compared to classical 1D stellar
evolution codes, many numerical innovations have been introduced to deal
with this complex problem.  First, the spectral discretization based on
spherical harmonics and Chebyshev polynomials is used to represent the
2D axisymmetric fields. A nonlinear mapping maps the spheroidal star and
allows a smooth spectral representation of the fields. The properties
of Picard and Newton iterations for solving the nonlinear partial
differential equations of the problem are discussed. It turns out that the
Picard scheme is efficient on the computation of the simple polytropic
stars, but Newton algorithm is unsurpassed when stellar models include
complex microphysics. Finally, we discuss the numerical efficiency of our
solver of Newton iterations. This linear solver combines the iterative
Conjugate Gradient Squared algorithm together with an LU-factorization
serving as a preconditionner of the Jacobian matrix.
\end{abstract}


\end{frontmatter}

%% \linenumbers

%% main text
\section{Introduction}

The recent progress of observational stellar astrophysics in
spectroscopy, spectropolarimetry or interferometry have called for more
realistic models of stars, with a focus on the effects of rotation.
Without rotation stars may be modeled as spherical 'balls' with a detailed
microphysics: equation of state, opacities or nuclear reaction rates
have been the subject of intense research over the past fifty years
\cite[e.g.][]{maeder09,KWW12}. In these one-dimensional models, the main
difficulty comes from the modeling of the averaged heat transport by
convection in the various parts of the star where hydrostatic equilibrium
is unstable. For stars burning hydrogen on the so-called main sequence,
these regions are a convective core when the stellar mass is larger than
1.3 solar mass - hereafter noted \msun\ - and a convective envelope when
the mass is less than 1.8~\msun.  One-dimensional models have been
designed and redesigned for more than fifty years now and are still
widely used (e.g. the code MESA started by \citealt{paxton_etal11}).
They have had great successes in depicting a now widely accepted view
of stellar evolution.

But, as alluded above, the more precise observations obtained with modern
instruments show details that are difficult to explain with
one-dimensional, spherically symmetric models. Most of these details are
related to fluid flows in the stars. We easily understand that it is
uneasy to model fluid flows in one dimension. The bulk effects of
rotation are the first victims of an imposed spherical symmetry. Current
one-dimensional codes, like MESA or CESAM \cite[][]{Morel97}, include a
modeling of rotation through its average effects: these are mainly the
centrifugal effect and radial differential rotation that mimic baroclinic
flows. As expected, these models show discrepancies when compared to
observational data. For instance, they have difficulties to reproduce the
abundances of elements at the surface of stars \cite[][]{brott_etal11}
or they simply cannot be used to interpret interferometric observations
of fast rotating stars \cite[e.g.][]{monnier_etal07}.

To overcome these difficulties, the natural step forward is to relax the
spherical symmetry in the modeling and to work with models owing two
dimensions of space at least. Thus, fluid flows can be computed more
realistically and rotational effects as well. The first step in this
direction is to elaborate two-dimensional axisymmetric models of stars.
The centrifugal distortion of the star can then be naturally included as
well as the global steady flows.

Attempts to build such models have begun in the sixties
\cite[][]{james64}, almost at the same time as 1D models. First steps
in the quest of 2D stellar models for fast rotating stars have been
marked by a series of works starting with the one of \cite{OM68} who
introduced the Self-Consistent Field (SCF) method\footnote{Briefly, this
method use's the formal solution of Poisson's equation in term of the
density distribution, i.e.

\[ \phi(\vx) = -G\int \frac{\rho(\vx')}{|\vx-\vx'|}d^3\vx'\]
which has the great advantage of including the boundary
conditions on $\phi$ at infinity. The potential is used to find
a new $\rho$ itself leading to a new potential.}\cite[see][for a
short historical review]{R06c}.  In a subsequent series of papers,
\cite{clem74,clem78,clem79,clem94} proposed another way of solving
Poisson's equation by using finite differences, while later on
\cite{EM85,EM91} introduced a linear mapping $r_i(\theta_k) = \zeta_i
R_s(\theta_k)$ such that the grid $r_i(\theta_k)$ automatically adjusts
to the shape of the star (here given by its colatitude dependent radius
$R(\theta)$). More recently, \cite{Roxburgh04,Roxburgh06} reconsidered
2D models of fast rotating stars for asteroseismic purposes, while
\cite{JMS04,JMS05} reconsidered similar models for interpreting the very
flattened shape of the Be star Achernar, as revealed by the first
precise interferometric observations of this star
\cite[][]{domiciano_etal03}. At the same time, \cite{JMS04,JMS05}
improved the SCF method. Recent results of \cite{macgregor_etal07}
presented SCF models with very high angular momentum showing stellar
models with very strongly distorted shapes compared to the sphere.
In an other line of research, \cite{deupree11} also computed 2D
models that he later used to interpret recent interferometric and
asteroseismic data obtained for the nearby fast rotating star Rasalhague
\cite[e.g.][]{deupree_etal12}. However, in all the foregoing work the
internal rotation of the star had to be specified (either as a solid
body rotation or as a given differential rotation). In real
isolated stars, differential rotation emerges from the baroclinic torque
and Reynolds stresses, the former being prominent in radiative zones
and the latter in convective regions. The first models that included
self-consistently the pressure, density, temperature distributions and
the associated baroclinic torque have been presented in \cite{ELR07}
and later, using the proper spheroidal geometry, in \cite{REL09,ELR13}.

The main difficulty was to find the appropriate algorithm that allowed
convergence of the iterations to the quasi-steady state of a
fast rotating star consistently with the mean flows that pervade
the whole star. Indeed, these flows face extremely large density
variations (typically eight orders of magnitude) making solutions prone to
numerical instabilities. In addition, heat transfer depends on the
strongly varying heat conductivity (controlled by the fluid opacity)
or on a vigorous turbulent convection. Even if thermal convection is
modeled by a smooth mean-field approach, the global rapid variations of
transport coefficients, especially near the surface, make the problem
thorny.

\begin{figure}\centering
\begin{minipage}[t]{.49\textwidth} \vfill \centering
\includegraphics[width=\linewidth,clip=true]{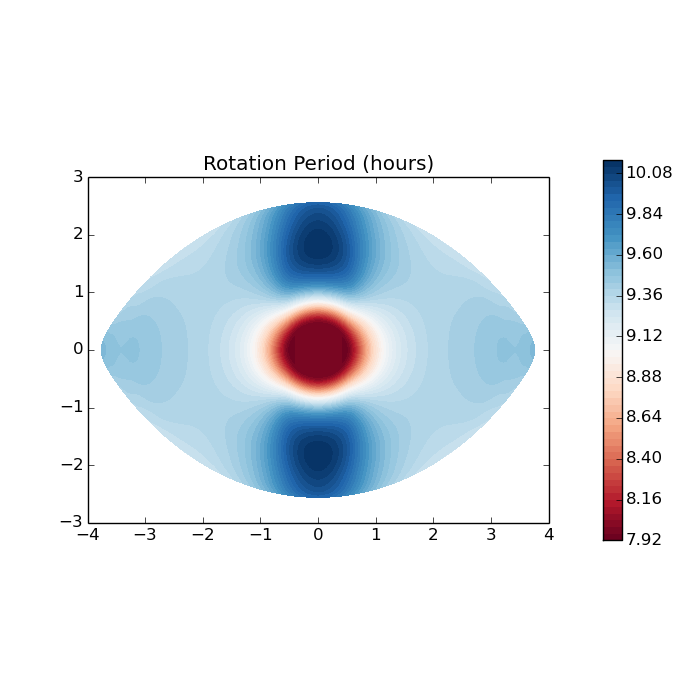}
\caption[]{Meridional cut of a 5\msun\ 2D ESTER (stellar) model showing
the internal differential rotation of the star. The equatorial velocity
is 95\% of the break-up velocity. Dark blue is for low angular velocity
and red for high angular velocity (core region). The side length scale unit is
the solar radius.
}
\label{rotdiff}
   \end{minipage} \hfill
   \begin{minipage}[t]{.49\textwidth} \vfill \centering
\includegraphics[width=\linewidth,clip=true]{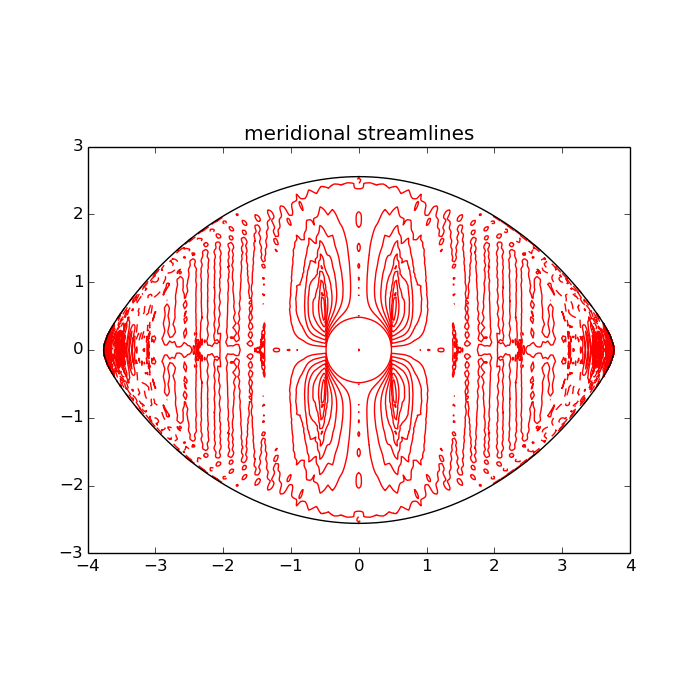}
\caption[]{Streamlines associated with the meridional circulation for
the same model and as in figure~\ref{rotdiff}.}
\label{mercirc}
   \end{minipage}

\end{figure}

The aim of this paper is to present to the readers the numerical side of
the solution that we have found to the modeling of fast rotating main
sequence stars as illustrated in Fig.~\ref{rotdiff} and
\ref{mercirc}. This solution is now used in the ESTER code, which is
freely available at http://ester-project.github.io/ester/. A detailed
discussion of the physical and astrophysical hypothesis of the ESTER
models may be found in \cite{ELR13} or \cite{REL13}. In the following,
we shall first present the set of equations to be solved (sect. 2)
and continue on presenting the mapping that is used to deal with the
spheroidal shape of the star (sect. 3). We then introduce our choice
of the discretization (spectral methods)  in section 4 and discuss the
choice of the algorithm (sect.  5). We finally illustrate the results with
examples showing the numerical efficiency of the ESTER code at computing
various stellar models (sect. 6). Conclusions and outlooks end the paper.

\section{Mathematical formulation}

\subsection{Equations of stellar structure}

Basically equations that are governing the structure of stars are those
governing a compressible self-gravitating fluid flow with heat sources.
Because of the very high temperatures of the central regions, heat sources
are coming from nuclear reactions. The (quasi) steady state is possible
because the star cools through radiation at its surface usually assumed
as a black body.  This is not a strict steady state because nuclear
reactions slowly change the chemical composition of the star leading
to higher and higher densities at the center. However, on the main
sequence, the time scale of hydrogen nuclear burning is sufficiently
long compared to all dynamical time scales so that the steady state is
a very good approximation. The present work restrict to this period of
the life of stars.

Let us now present the four partial differential equations that govern
this steady state.

\subsubsection{Gravitational potential}

Self-gravity is governed by Poisson's equation

\begin{equation}
\nabla^2\phi = 4\pi G\rho
\label{poisson}
\end{equation}
where $\phi$ is the gravitational potential, $\rho$ is the density and
$G$ the gravitational constant. This equation is completed by boundary
conditions that state that $\phi$ is regular at the center of the star
and vanishes at infinity.

\subsubsection{Dynamics}

Two equations govern the mean velocity field in the star. Indeed, we are
only interested in the average steady state of the star and therefore
turbulent flows are only represented by their mean-fields. Hence,
everywhere in the following, $\vv$ should be understood as a mean velocity
field. Moreover, no magnetic field is considered.
Mass conservation yields

\begin{equation}
\nabla\cdot(\rho\vv) = 0
\end{equation}
while the momentum conservation requires

\begin{equation}
\rho \vv\cdot\na\vv = -\na P -\rho\na\phi+\vF_v
\label{mom}
\end{equation}
in an inertial frame. In these equations, $P(\vr)$ is the mean pressure
and $\rho(\vr)$ the mean density. The viscous force $\vF_v$ includes,
whenever necessary, the turbulent Reynolds stresses. It is taken
into account in order to remove the degeneracy of the inviscid problem
\cite[e.g.][and below]{ELR13}. For later use, we recall that the
equatorial angular velocity is bounded by the keplerian angular velocity
at equator, namely

\beq \Omega_k = \sqrt{\frac{GM}{R^3_{\rm eq}(\Omega_k)}} \eeq
We shall call this quantity the critical angular velocity. The
dependance $R_{\rm eq}(\Omega_k)$ underlines that the equatorial
radius depends on the angular velocity at equator, hence the
determination of the critical angular velocity needs the solution of an
involved nonlinear problem.

\subsubsection{The temperature field}\label{sectemp}

The equation of energy or of entropy $S$ gives the temperature field. We
write it as

\begin{equation} \rho T \vv\cdot\nabla S = -\na\cdot\vF +
\eps_*\label{eq_temp}\end{equation}
where the heat flux is

\beq \vF = -\khi\na T +\vF_{\rm conv}\eeq
namely the sum of the radiative flux and the convective flux when thermal
convection sets in. In these equations, $\eps_*\equiv\eps_*(\rho,T)$
is the heat power generated by nuclear reactions per unit volume and
$\chi\equiv\chi(\rho,T)$ is the heat conductivity.

Convective zone boundaries are determined by the Schwarzschild
criterion. Namely, we associate convectively stable regions with those
verifying 

\[ -\vg_{\rm eff}\cdot\na S >0 \]
Note that our models have a chemically homogeneous core and envelope.
Thus Ledoux criterion does not apply and our model do not deal with
semi-convection.

The convective flux needs being modeled by a mean-field
approach. Presently, no general model exist. One dimensional models have
shown that the convective core of massive or intermediate-mass stars
is almost isentropic. We therefore assume perfect isentropy in stellar
convective cores. Elsewhere, and in particular in stellar envelopes, convection
is not efficient enough to impose isentropy everywhere. In 1D models,
one uses the mixing length theory \cite[][]{maeder09}, but this approach
has not yet been generalized to a 2D set-up. Thus, we shall ignore at
the moment stars with convective envelopes. This restricts the model
applications to main-sequence stars of high or intermediate mass (the
so-called early-type stars whose mass is typically larger than
2~\msun). In these stars surface convection may exist but it is
inefficient and does not affect the stellar structure.

\subsection{Boundary conditions}

The foregoing partial differential equations need to be completed
by boundary conditions. At the star's center we just need to impose
the regularity of the fields. At the stellar surface the situation is
more complicated.  First, we need to define the stellar surface.

\subsubsection{The stellar surface}

As in 1D models we define the stellar surface as given by a fixed
optical depth. We recall that the optical depth is a non-dimensional
quantity, usually called $\tau$, defined by

\[ \tau(z,\nu) = \int^\infty_z \alpha_\nu(z')dz'\]
Here, $\alpha_\nu(z')$ is the absorption coefficient at the altitude $z'$
and at the (electromagnetic) frequency $\nu$. Usually, a grey atmosphere
model is adopted to avoid the frequency dependence. Typically, the
surface of the star is defined as the surface where the grey optical
depth is 2/3 \cite[][]{HK94}. With a simple model of the atmosphere
\cite[e.g.][]{ELR13}, this surface is replaced by an isobar whose pressure
is determined at the pole of the star (see below).

\subsubsection{The pressure and temperature boundary conditions}

The pressure is determined through its gradient in the momentum
equation \eq{mom}. Hence, it is known up to a constant. This constant actually
fixes the extension of the star and thus its radius. To be consistent with
spherically symmetric models we fix the pressure constant by assuming
that the polar pressure is given by

\beq P_s= P_{\rm pole} = \tau_s\frac{g_{\rm pole}}{\kappa_{\rm pole}},\eeqn{polpres}
where $\tau_s$ is the chosen optical depth (usually 2/3), $g_{\rm pole}$
is the polar gravity and $\kappa_{\rm pole}$ is the polar opacity
(opacity is related to the absorption coefficient by
$\kappa=\alpha/\rho$). Relation \eq{polpres} is derived from a simplified
hydrostatic model of the atmosphere of stars \cite[e.g.][]{KWW12}.

Once the polar pressure is defined by \eq{polpres}, the associated
isobar is taken as the surface of the stellar model. This surface coincides
with the photospheric surface of the star at the pole only and is below the
photospheric surface elsewhere. To take into account the fluid above
this isobar, we model this fluid layer by a locally polytropic
atmosphere. Namely, we say that in this atmosphere

\beq P= P_{\rm pole}(1-z)^{n+1}\eeqn{poly_atm}
where $n$ is the polytropic index and $z$ a scaled height in the
atmosphere. With \eq{poly_atm}, we can set the true surface at $z=0$ at
the pole and at $z(\theta)$ elsewhere. At the true surface, where the
optical depth equals $\tau_s$ the pressure is

\[ \tau_s\frac{g_{\rm eff}(\theta)}{\kappa(\theta)}\]
and $z(\theta)$ is such that

\[ 1-z(\theta) = \lp\frac{g_{\rm eff}(\theta)\kappa_{\rm pole}}{g_{\rm
pole}\kappa(\theta)}\rp^{1/(n+1)}\; .\]
Using the fact that the temperature varies like $1-z$ in a polytropic
atmosphere, and that the temperature equals the effective temperature
at optical depth $\tau_s$, we find that
the temperature must verify

\beq T=T_s(\theta) = T_{\rm eff}(\theta)\lp\frac{g_{\rm
pole}\kappa(\theta)}{g_{\rm eff}(\theta)\kappa_{\rm pole}}\rp^{1/(n+1)}
\eeq
on the isobar $P=P_{\rm pole}$. We note that according to this definition,
$T_s($pole$)$ is also the polar effective temperature of the stellar
model. Hence, along with the regularity of the temperature field at the
star's center, the condition

\[ T=T_s(\theta)\qquad {\rm at\;the\; surface}\]
fully defines the temperature field inside the stellar model
\cite[][]{ELR13}. For applications below, we set the polytropic index to
$n=3$, a value adapted to a radiative layer.

\subsubsection{Velocity boundary conditions}

The natural boundary conditions for the velocity field at the surface of
the star are the stress-free conditions. Indeed, the neglected layers
actually impose a negligible stress on the surface (this simplified
view may however be challenged in real stars by a combination of winds
and magnetic fields). These conditions read:

\beq \vv\cdot \vn =0 \quad {\rm and}\quad ([\sigma]\vn)\wedge\vn =\vzero
\eeqn{bcvel}
where $[\sigma]$ is the stress tensor and $\vn$ the outer normal of the
stellar surface. Actually, we have to take into account the additional
constraint that the equatorial velocity is given as needed to fix the
rotational velocity of the star, namely

\beq v_\varphi(r=R_{\rm eq},\theta=\pi/2) = V_{\rm eq} \eeqn{veq}

\subsubsection{Dealing with viscosity}

The boundary conditions on the velocity \eq{bcvel} are numerically costly
because in stellar conditions thin boundary (Ekman) layers develop at
the surface. {However, simplifying these conditions by just
neglecting viscosity altogether is not possible. Indeed, it is
well-known that the inviscid problem is degenerate \cite[e.g.][]{R06}.
Using \eq{mom}, imposing axisymmetry and neglecting the viscous force,
we get the so-called thermal wind equation \cite[][]{green69}, which
here reads

\begin{equation}
\label{eq:vort_inv}
s\frac{\partial\Omega^2}{\partial z}=\ephi\cdot\frac{\na
p\times\na\rho}{\rho^2} \;,
\end{equation}
where $\Omega\equiv\Omega(s,z)$ is the local angular velocity and
$(s,z,\varphi)$ are the cylindrical coordinates. We
easily see that this equation is invariant in the transformation
$\Omega^2\tv\Omega^2+\Omega^2_g(s)$ for any $\Omega_g(s)$. The condition
imposing the equatorial velocity \eq{veq} is not sufficient to lift the
degeneracy. Actually, the same problem arises when one considers the steady
baroclinic flows in a rotating frame \cite[e.g.][]{R06}. Hence,
viscosity plays a fundamental role in the determination of the
differential rotation of a star (especially in the radiative regions),
but it brings new very thin boundary layers that make the problem more difficult
numerically. However, using the results of \cite{R06}, \cite{ELR13}
have devised a new (nonlinear) boundary condition for $\Omega(s,z)$
that couples this quantity with the stream function of the meridian
circulation, and which avoids the computation of the Ekman layers.
This boundary condition reads

\begin{equation}
\label{eq:bl}
\mu
s^2\mathbf{\hat\xi}\cdot\na\Omega+\psi\mathbf{\hat\tau}\cdot\na(s^2\Omega)=0
\qquad\mbox{on the surface}\;.
\end{equation}
where $\mu$ is the dynamical viscosity and $\mathbf{\hat\xi}$ is a unit
vector perpendicular to the surface while $\mathbf{\hat\tau}$ is a unit
vector tangent to it. $\psi$ is the stream function of the meridional
flow $\vu$, such that $\rho\vu=\na\times(\psi\hat\varphi)$, $\hat\varphi$
being the azimuthal unit vector. The derivation of \eq{eq:bl} is quite
tedious and we refer the reader to \cite{ELR13} for the details.}
Thus doing, viscosity (if small enough!) can be taken into account for
the determination of the azimuthal velocity $v_\varphi(r,\theta)$ and
the associated angular velocity $\Omega(r,\theta)$, without including
explicitly the viscous force in the meridional part of the momentum
equation. Hence, the meridional part of the momentum equation \eq{mom}
reduces to Euler's equation

\begin{equation}
\rho s\Omega^2\es=\na p+\rho\na\phi
\label{baroc}
\end{equation}
taking into account the axisymmetry of the fields and the vanishing of
the meridian circulation with viscosity \cite[e.g.][]{busse81,R06}.
Although the variables are all strongly coupled, we may consider
\eq{baroc} as the equation determining the pressure field.

The $\varphi$-component of \eq{mom}, controls the balance of the flux of
angular momentum and determines the meridional velocity (along with mass
conservation) forced by the viscous stress generated by the (previously
derived) differential rotation. It reads

\begin{equation}
\label{eq:angular_mom}
\na\cdot{(\rho s^2\Omega\vu)}=\na\cdot (\mu s^2\na\Omega)
\end{equation}
where $\vu$ is the meridional circulation and $\mu$ the dynamical
viscosity. For simplicity we here assume that shear stresses are
represented by a mere (newtonian) viscous force, but in principle
more elaborated Reynolds stress models can be used. Equation
\eq{eq:angular_mom} may be understood as the one used to derive the stream
function $\psi$ of the meridional circulation. From \eq{eq:angular_mom},
we now see that the profile of viscosity $\mu$ influences the shape
of the meridional circulation and through \eq{eq:bl}, the differential
rotation. However, the amplitude of the viscosity (in the limit of vanishing
Ekman numbers) has no importance for the shape of the flow field (see
below).

The foregoing treatment of the effects of viscosity has the merit of
eliminating the small scale induced by the Ekman layer, while removing
the degeneracy of the inviscid limit. There is however a price to pay:
the elimination of the viscous effects in the meridional component of
the momentum equation \eq{baroc} also eliminates the (viscous) Stewartson
layer that naturally arises on the tangent cylinder circumventing the
convective core \cite[e.g.][]{ELR13}. More work is needed to find a
way to take into account such a dynamical feature, which presumably
plays a role in the transport of chemical elements, while preserving
the numerical stability of the solutions.

\subsection{Microphysics}

Beside the foregoing partial differential equations and their associated
boundary conditions, one needs to specify the equations of state of the
fluid, the dependence of the radiative conductivity with temperature and
density, and the power of nuclear reactions. As far as the equation of
state and the opacity are concerned (the opacity controls the radiative
conductivity), we use the OPAL tables of \cite{RSI96}, which give the
relations

\beq P\equiv P(\rho,T) \eeq
\beq \chi \equiv \chi(\rho,T)\eeq
as well as other thermodynamics quantities.

For the nuclear heat power, we use an analytic formula that
accounts for the heat generation by hydrogen fusion either by the
pp-chains or by the CNO cycle (each dominating in some range of
temperatures). For that we set

\beq \eps_* = \eps_*^{\rm pp} + \eps_*^{\rm CNO} \eeq
with

\beq \eps_*^{\rm pp} = \eps_0^{\rm
pp}X^2\rho^2T_9^{-2/3}\exp\lp-\frac{A_{\rm pp}}{T^{1/3}} \rp\; .
\eeq
This expression is also used by the CESAM code \cite[][]{ML08}.
Similarly,

\beq
\eps_*^{\rm CNO}(\rho,T)=\eps_0^{\rm CNO}XX_{\rm CNO}\rho^2T_9^{-2/3}
\exp\lp-\frac{A_{\rm CNO}}{T^{1/3}}\rp
\times\lp1+0.027T_9^{1/3}-0.778T_9^{2/3}-0.149T_9\rp, \eeq
using the expression given in \cite{KW90}\footnote{\cite{KWW12} give a
more recent expression of the CNO-cycle power but the difference with
the ``old" expression is negligible for our purpose.}. In these expressions
$T_9=T/10^9$, $X$ is the hydrogen mass fraction, $X_{\rm CNO}\simeq Z/2$
is the mass fraction of CNO elements assumed to be a solar mixture of
metallicity $Z$ \cite[e.g.][]{maeder09}. The constants are taken as:

\[ \eps_0^{\rm 
pp} = 8.24\times10^4 \;{\rm cgs}, \qquad \eps_0^{\rm CNO} =
8.67\times10^{25} \;{\rm
cgs}, \qquad A_{\rm pp} = 3600., \quad A_{\rm CNO}= 1.5228\times10^4\]
but detailed nuclear reaction rates can be obtained from the NACRE
compilation \cite[][]{angulo_etal99,xu_etal13}.

\subsection{Integral constraints}

The foregoing equations are completed by one integral
constraint, namely the one that specifies the mass of the star

\beq M= \intvol\rho dV\; .
\eeq
If needed, the total angular momentum

\begin{equation} \vL = \intvol \rho \vr\times \vv dV\label{angmom}
\end{equation}
can also be imposed, but in such a case this constraint replaces the fixed
equatorial velocity \eq{veq}.

\subsection{Scaled equations}

In order to solve the foregoing set of equations, we first scale
the equations so as to use, when possible, non-dimensional variables.

 We choose
to scale pressure, density and temperature by their central values and
other quantities as follows:

\begin{center}\parbox{10cm}{

Length scale $\equiv$ polar radius \dotfill $R$

Pressure scale $\equiv$ central pressure \dotfill $P_c$

Density scale $\equiv$ central density \dotfill $\rho_c$

Temperature scale $\equiv$ central temperature \dotfill $T_c$

Gravitational potential scale \dotfill  $P_c/\rho_c$

Angular velocity scale \dotfill $\frac{1}{R}\sqrt{\frac{P_c}{\rho_c}}$

Meridional velocity scale \dotfill $E\sqrt{\frac{P_c}{\rho_c}}$
}
\end{center}\bigskip

\noi where $E$ is the Ekman number defined as

\begin{equation} E = \frac{\mu_c}{\rho_c \Omega_0 R^2} \with
\Omega_0=\sqrt{\frac{P_c}{R^2\rho_c}}\; .\end{equation}

With these scalings, Poisson's equation now reads:

\begin{equation} \nabla^2\phi = \pi_c \rho \with \pi_c = \frac{4\pi
G\rho_c^2}{P_c}\label{poissonnd}\end{equation}
{ The scaled equation of angular momentum reads

\begin{equation}
\label{eq:angular_momnd}
\na\cdot{(\rho s^2\Omega\vu)}=\na\cdot (\mu s^2\na\Omega)
\end{equation}
where $\mu\equiv\mu(r,\theta)$ is the normalized viscosity profile. As a
first step, and in the following, we simply take $\mu=1$.

With the foregoing scalings, we note that the meridional circulation is
very small compared to the differential rotation. Indeed, from
\eq{eq:angular_momnd} we see that $\|\vu\|\sim\Omega\sim1$, but
dimensional velocities are such that $\|\vv_{\rm merid}\|/V_\varphi \sim
E$, while in fast rotating stars $E\infapp10^{-8}$ \cite[e.g.][]{ELR13}.
Hence, the scaled  meridional circulation together with boundary
conditions \eq{eq:bl}, which now reads

\begin{equation}
\label{bc_adim}
s^2\mathbf{\hat\xi}\cdot\na\Omega+\psi\mathbf{\hat\tau}\cdot\na(s^2\Omega)=0
\qquad\mbox{on the surface}
\end{equation}
with scaled variables, allows us to set $E=0$ without introducing an
undetermined function in the solution.

Another benefit of taking the limit $E=0$ is that it eliminates heat
advection in the energy-entropy equation solved in radiative regions. Such
limit is indeed equivalent to setting the Prandtl number to zero which
stresses that heat is more efficiently transported by diffusion than by
advection in a radiative region \cite[see a recent discussion of the
dynamics of rotating radiative regions in][]{R06b}. Hence, the energy
equation can be simplified as

\begin{equation} \nabla^2 T + \na\ln\chi\cdot\na T +
\Lambda\frac{\eps_*}{\chi_*} = 0 \with \Lambda = \frac{\rho_c R^2}{T_c}
\label{temp_scaled}
\end{equation}
Note that $\Lambda$ is a dimensional constant since $\eps_*$ and $\chi_*$
are of different dimensions.

While \eq{temp_scaled} is solved to give the temperature field in
radiative zones, convective regions need a specific treatment.
Presently, our solutions only handle convective cores where we impose

\beq \mathbf{\na} S=\vzero \eeqn{eqentropy}
with $S$ being the entropy. Such an equation assumes that convection is
extremely efficient so as to impose a constant entropy everywhere in the
convective region. This is actually the case in stellar convective cores
according to the mixing-length model \cite[][]{maeder09}. In convective
envelopes, like that of the Sun, the efficiency of convection decreases
very much near the stellar surface and \eq{eqentropy} is no longer
appropriate. A new approach is needed, but it is beyond the scope of the
present work. Thus, as mention in the introduction, the present
numerical solutions only apply to main-sequence early-type stars, since
these stars own a convective core (due to a powerful nuclear heating),
but no or a very small convective envelope.

To finish with the scaled equation, we note that the momentum equation
\eq{baroc} and the vorticity equations \eq{eq:vort_inv} are unchanged
(except that they use scaled variables).  Mass conservation also remains
the same

\begin{equation} \na\cdot(\rho\vu) =0\label{massc}\end{equation}
}

\section{The mapping}\label{sectmap}

One of the difficulties of this problem is the {\it a priori} unknown shape of
the star. Hence, we devised a method where the grid evolves with
iterations so as to always fit the surface of the star where boundary
conditions are applied. 

\begin{figure}[t]
\centerline{
\includegraphics[width=0.9\linewidth,clip=true]{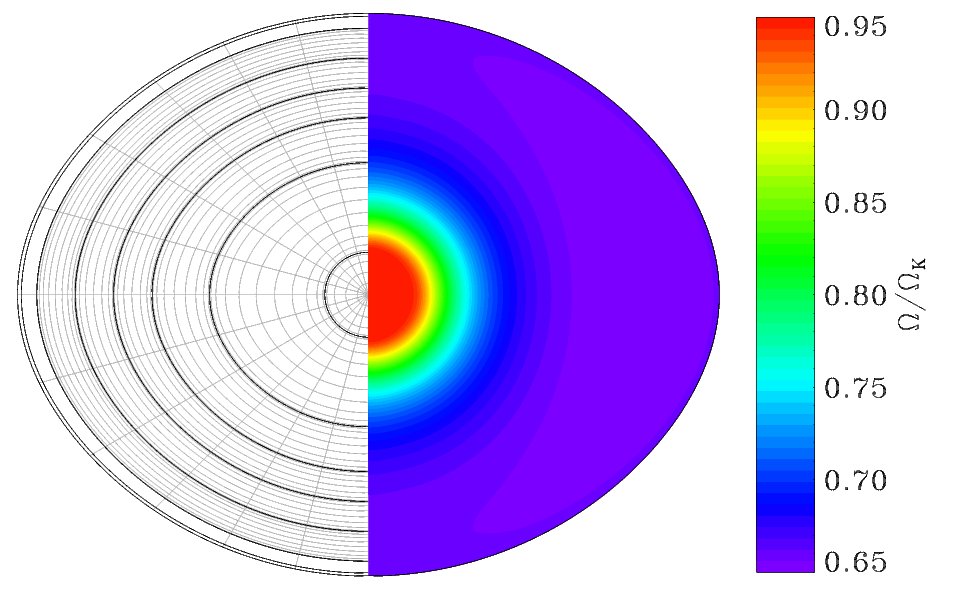}}
\caption[]{Left: plot of the computational grid inside the star. The
solid thick lines show the frontiers of the domains. The outer grid is
not represented. Right: just for illustration, the associated differential
rotation of the computed stellar model with parameters M=3~\msun\ and
$\Omega/\Omega_c=0.7$ leading to a flatness of 20\%. The chemical
composition is homogeneous and solar. (Credit Daniel Reese)}
\label{grid}
\end{figure}

\subsection{An adapted system of coordinates}

We therefore construct a mapping which connects the usual spherical
coordinates $(r,\theta,\varphi)$ to new spheroidal coordinates
$(\zeta,\theta'\varphi')$ such that the surface $\zeta=1$ describes the
surface of the star. Therefore the mapping reads:

\begin{equation}
\left\{
\begin{array}{l}
r=r(\zeta,\theta')\\
\theta=\theta'\\
\varphi=\varphi'
\end{array}
\right.
\label{themap}
\end{equation}
The problem reduces to that of finding a suitable form for the function
$r(\zeta,\theta)$.

However, the mapping needs to be a little more sophisticated. Indeed, we
also need to solve Poisson's equation outside the star since the
gravitational field is not simple in this domain. Thus, the
computational domain should be at least divided into two parts: the
star and the surrounding vacuum. But our choice of a spectral method
(see below) makes the use of a multidomain method attractive also inside
the star. It indeed turns out that a spectral solver deals more
easily with the dramatic variations of density between the center and the
surface of the star when these density variations are distributed over
several domains. Thus, we split the star into $n$-spheroidal shells.
More precisely, we distribute the domains so that the pressure ratio
(between the upper and lower boundaries of the domain) is almost the same in
every domains. Thus, we easily deal with the changing pressure scale
height within the star. The domain boundaries can also be attached to
discontinuities that arise from the physics of the star (chemical
barriers, jumps in thermal gradients, etc).

Hence, we divide the domain $\calD$ into $n$ subdomains $\{\calD_i\}$,
the frontiers of which are given by a series of functions $\{R_i(\theta),
i=0,n-1\}$ such that the domain $\calD_i$ is bounded by the spheroids

\[ r=R_i(\theta) \andet r=R_{i+1}(\theta)\]
where $R_0(\theta)=0$ is the center of the star and
$R_n(\theta)=R(\theta)$ is the stellar surface. We also use an
external domain $\mathcal{D}_{ex}$ that extends from the outer boundary
$R_n(\theta)$ to infinity. In this latter domain, we only solve
Poisson's equation. This additional domain allows us to impose conveniently the
vanishing of the gravitational potential at infinity. In Fig.~\ref{grid}
we give an example of the grid structure that is used inside the star
together with the differential rotation associated with the solution.

\subsection{The details of the mapping}

Inspired by the work of \cite{BGM98}, we choose the following form
of the mapping in the stellar domains $\calD_i$

\begin{equation}
\label{eq:map}
r(\zeta,\theta)\equiv r_i(\zeta,\theta) =
a_i\xi\Delta\eta_i+R_i(\theta)+A_i(\xi)(\Delta
R_i(\theta)-a_i\Delta\eta_i) 
\qquad \mbox{for $\zeta\in[\eta_i,\eta_{i+1}]$}
\end{equation}
where the $a_i's$ are constants and where we have defined:

\begin{eqnarray*}
\eta_i&=&R_i(\theta=0)\\
\Delta\eta_i&=&\eta_{i+1}-\eta_i\\
\Delta R_i(\theta)&=&R_{i+1}(\theta)-R_{i}(\theta)\\
\xi&=&\displaystyle\frac{\zeta-\eta_i}{\Delta\eta_i} \quad {\rm so\;
that}\quad 0\leq\xi\leq 1\quad {\rm when}\quad\zeta\in[\eta_i,\eta_{i+1}]
\end{eqnarray*}
Since the polar radius is chosen as the unit length, $R(0)\equiv
R_n(0)=1$ and $\eta_n=1$. Hence, the star is scanned when $\zeta\in[0,1]$.
Note that the $\xi$ variable is a local radial variable specific to each
domain. The mapping in the outer empty domain surrounding the star will
be discussed below.

The $A_i(\xi)$ functions verify

\[ A_i(0) = 0 \andet A_i(1) = 1\]
so that

\[ r(\eta_i,\theta) = R_i(\theta) \]
but are otherwise arbitrary.

A simple choice of the $A_i$-functions would be $A_i(\xi)=\xi$ so
that the mapping is linear, namely

\beq r(\zeta,\theta) = R_i(\theta)+\xi\Delta R_i(\theta) \for
\xi\in[0,1] \andet \zeta=\eta_i+\xi\Delta\eta_i
\eeqn{lin_map}
If the star is described by a single domain, then

\beq r(\zeta,\theta) = \xi R(\theta)\eeq
This is the mapping that was chosen by \cite{EM85,EM91}. However, as
noticed by \cite{EM85} and \cite{BGM98}, this linear mapping requires
a special treatment of the center. This is why \cite{BGM98} suggested
to use a nonlinear mapping based on a higher order polynomial for the
$A_i$-functions. Hence, following this latter work, we choose

\begin{eqnarray}
\label{eq:map_bonazzola}
A_0(\xi)=(5\xi^3-3\xi^5)/2&& \\
A_i(\xi)=3\xi^2-2\xi^3 &\qquad& \mbox{for $i=1,\ldots,n-1$}
\end{eqnarray}
This choice is such that near the center of the star, the
$(\zeta,\theta,\varphi)$ coordinates reduce to the spherical
coordinates $(r,\theta,\varphi)$. This property allows us to use the
properties of polynomial regularity of functions expanded over spherical
harmonics. Central boundary conditions can thus be imposed without
numerical difficulties.

\bigskip
The $a_i$ constants are arbitrary and should be optimized for the problem
at hands.  Here, they are chosen so that $r(\zeta,\theta)$ is an
increasing function of $\zeta$ and so as to avoid a singular mapping.
The Jacobian of the mapping is

\[ J=\left|\begin{array}{cc}
\dzeta{r} & \frac{\partial r}{\partial\theta'}\\
\dzeta{\theta} & \frac{\partial\theta}{\partial\theta'}
\end{array}\right| = \dzeta{r} \equiv r_\zeta\]
We first observe that:

\begin{equation}
A'_i(\xi) = 6\xi(1-\xi) \andet A'_0 = \frac{15}{2}\xi^2(1-\xi^2)
\end{equation}
Hence, $A'_i(0)=A'_i(1)=0$, and

\[ J=a_i+A_i'(\xi)\left(\frac{\Delta
R_i(\theta)}{\Delta\eta_i}-a_i\rp\]
in domain $\calD_i$. From the foregoing expressions of $A'_i$, we
note that $A'_i(\xi)\geq0$ in each domain. On the other hand, the
thickness of each domain, namely $\Delta R_i(\theta)$, can be chosen
as an increasing function of $\theta$ from pole to equator. This is
indeed always possible as rotating stars are oblate, namely with an
equatorial radius larger than the polar radius. Hence, we can choose the
$R_i(\theta)$ such that 

\[ \Delta R_i(\theta)\geq \Delta\eta_i, \qquad
\forall\;\theta\in[0,\pi]\; .\]
Thus, if we choose

\[ a_i=1\]
we are insured that

\[ J(\zeta,\theta) \geq 1 \qquad \forall \zeta,\theta \in {\rm the\;
star}
\]
We thus also satisfy the constraint $r_\zeta>0$, which insures that $r$ is a
monotonically increasing function of $\zeta$. In practice we shall set
$R_i(\theta)$ surfaces to isobars.

With the above considerations, we adopt the following mapping

\begin{equation}\label{themapp}
r(\zeta,\theta) \equiv r_i(\zeta,\theta) = \xi\Delta\eta_i + R_i(\theta) + A_i(\xi)(\Delta
R_i(\theta)-\Delta\eta_i), \with \xi\in[0,1]
\end{equation}
for the domain $\calD_i$. From \eq{themapp} we find

\beqa
r_\zeta &=& 1+A_i'(\xi)\left(\frac{\Delta
R_i(\theta)}{\Delta\eta_i}-1\right) \\
r_\theta &=& R'_i(\theta) +A_i(\xi)\Delta R'_i(\theta) \\ 
r_{\zeta\theta} &=& A_i'(\xi)\frac{\Delta R_i(\theta)}{\Delta\eta_i} \\
r_{\zeta\zeta} &=& \frac{A_i''(\xi)}{\Delta\eta_i}\left(\frac{\Delta
R_i(\theta)}{\Delta\eta_i}-1\right)
\eeqa
where the primes indicate derivation.

The use of the $(\zeta,\theta,\varphi)$ coordinates leads to new
expressions of differential operators. We give a short account of their
new form together with the metric tensor and the needed tensorial
quantities in the appendix of the paper.

Finally, we should give the mapping in the outer empty domain surrounding
the star. There, we choose the following expression:

\beq r(\zeta,\theta)\equiv r_{\rm ex}(\zeta,\theta) = \zeta -1
+R_i(\theta) \with \zeta\in[1,+\infty[
\eeq
namely a linear mapping that smoothly continues the inner one in the
last domain. The grid points are of course distributed using another
mapping that connects the infinite domain to a finite one, namely

\[ \zeta=\frac{1}{1-\xi} \with \xi\in[0,1]\]

\subsection{Interface conditions between domains}

At the domain boundaries we need writing continuity conditions that link
the fields in each domain. The main issue is that the mapping given by
\eq{themapp} has discontinuities in some of its derivatives. One easily
finds that

\[ r_\zeta(\eta_i) = 1, \qquad r_\theta(\eta_i) = R'_i(\theta), \qquad
r_{\zeta\theta}(\eta_i)=0\]
so that all these quantities are continuous at the interfaces of the
domains. On the contrary $r_{\zeta\zeta}$ is not continuous. We summarize
these properties in Tab.~\ref{cont_map} along with those of the linear
mapping \eq{lin_map}. Although simpler, the linear mapping has more
discontinuities between domains.

\begin{table}
\begin{center}
\begin{tabular}{c|c|c}
& \multicolumn{2}{c}{Continuity between subdomains} \\
& Our mapping & The linear mapping \\
\hline
$r$ & Yes & Yes\\
$r_\zeta$ & Yes  & No\\
$r_{\zeta\zeta}$ & No & Yes \\
$r_\theta$ & Yes & Yes \\
$r_{\theta\theta}$ & Yes & Yes \\
$r_{\zeta\theta}$ & Yes & No \\
\end{tabular}
\end{center}
\caption[]{Continuity of the various derivatives at the
interface between domains for the linear mapping \eq{lin_map} and our
mapping.} \label{cont_map}
\end{table}

In the case of a scalar field $\phi(r,\theta)$, if $\phi$ is continuous
between subdomains, we simply need to impose

\begin{equation}
\phi^{(+)}=\phi^{(-)}
\end{equation}
where $(+)$ and $(-)$ represent each side of the domains interfaces. If
in addition we want $\phi$ to be derivable across the boundary, we have to
write a condition on its normal derivative $\vect{\hat n}\cdot\nabla\phi$
(and not on $\frac{\partial\phi}{\partial\zeta}$), namely,

\begin{equation}
\label{eq:map_cond_der}
\vect{\hat n}\cdot\nabla^{(+)}\phi^{(+)}=\vect{\hat
n}\cdot\nabla^{(-)}\phi^{(-)}
\end{equation}
Noting that

\begin{equation}
\vect{\hat n}\cdot\nabla\phi =
\sqrt{1+\frac{r_\theta^2}{r^2}}\left(\frac{1}{r_\zeta}\frac{\partial\phi}{\partial\zeta}-
\frac{r_\theta}{r^2+r_\theta^2}\frac{\partial\phi}{\partial\theta}\right)
\end{equation}
and that $r$, $r_\theta$ and $\partial_\theta\phi$ are continuous
across the interface, which is a $\zeta=\cst$ surface, then condition
(\ref{eq:map_cond_der}) becomes

\begin{equation}
\frac{1}{r_\zeta^{(+)}}\left(\frac{\partial\phi}{\partial\zeta}\right)^{(+)}=
\frac{1}{r_\zeta^{(-)}}\left(\frac{\partial\phi}{\partial\zeta}\right)^{(-)}
\end{equation}
that is equivalent to saying
$\displaystyle\left(\frac{\partial\phi}{\partial r}\right)^{(+)}=
\left(\frac{\partial\phi}{\partial r}\right)^{(-)}$. Since our mapping
ensures that
$r_\zeta^{(+)} = r_\zeta^{(-)}$, we also have

\beq
\left(\frac{\partial\phi}{\partial\zeta}\right)^{(+)}=
\left(\frac{\partial\phi}{\partial\zeta}\right)^{(-)}
\eeq

In the case of a continuously differentiable vector field, the conditions
of continuity of the field are simply those of the components, namely
\begin{equation}
\begin{array}{c}
\displaystyle  {V^\zeta}^{(+)}= {V^\zeta}^{(-)} \\
\displaystyle {V^\theta}^{(+)}={V^\theta}^{(-)} \\
\displaystyle {V^\varphi}^{(+)}={V^\varphi}^{(-)} 
\end{array}
\end{equation}
while the continuity of the normal derivative, namely $\vect{\hat
n}\cdot\nabla\vect{V}$, leads to

\begin{equation}
\begin{array}{c}
\displaystyle \left(\frac{\partial
V^{\zeta}}{\partial\zeta}\right)^{(+)}
+\frac{1}{r_\zeta^{(+)}}\left(r_{\zeta\zeta}^{(+)}{V^\zeta}^{(+)}+r_{\zeta\theta}^{(+)}{V^\theta}^{(+)}\right)=
\left(\frac{\partial V^{\zeta}}{\partial\zeta}\right)^{(-)}
+\frac{1}{r_\zeta^{(-)}}\left(r_{\zeta\zeta}^{(-)}{V^\zeta}^{(-)}+r_{\zeta\theta}^{(-)}{V^\theta}^{(-)}\right)
\\
\displaystyle \frac{1}{r_\zeta^{(+)}}\left(\frac{\partial
V^{\theta}}{\partial\zeta}\right)^{(+)}= 
\frac{1}{r_\zeta^{(-)}}\left(\frac{\partial
V^{\theta}}{\partial\zeta}\right)^{(-)} \\
\displaystyle \frac{1}{r_\zeta^{(+)}}\left(\frac{\partial
V^{\varphi}}{\partial\zeta}\right)^{(+)}= 
\frac{1}{r_\zeta^{(-)}}\left(\frac{\partial
V^{\varphi}}{\partial\zeta}\right)^{(-)}
\end{array}
\end{equation}
which reduces to

\begin{equation}
\begin{array}{c}
\displaystyle \left(\frac{\partial
V^{\zeta}}{\partial\zeta}\right)^{(+)}
+\frac{r_{\zeta\zeta}^{(+)}}{r_\zeta^{(+)}}{V^\zeta}^{(+)}=
\left(\frac{\partial V^{\zeta}}{\partial\zeta}\right)^{(-)}
+\frac{r_{\zeta\zeta}^{(-)}}{r_\zeta^{(-)}}{V^\zeta}^{(-)}
\\
\displaystyle \left(\frac{\partial V^{\theta}}{\partial\zeta}\right)^{(+)}= 
\left(\frac{\partial V^{\theta}}{\partial\zeta}\right)^{(-)} \\
\displaystyle \left(\frac{\partial V^{\varphi}}{\partial\zeta}\right)^{(+)}= 
\left(\frac{\partial V^{\varphi}}{\partial\zeta}\right)^{(-)}
\end{array}
\end{equation}
for our mapping. These conditions are equivalent to saying that the viscous
stress is continuous across the interface in case $\vV$ is a velocity
field. Note that this condition would be more complex with a linear
mapping because $r_\zeta$ is not continuous in this latter case.

\section{The discretization}

As far as discretization of the differential operators is concerned, we
chose a spectral grid in each domain. We thus use the Gauss-Lobatto grid
for the radial coordinate $\xi$ and the spherical harmonic expansion for
the horizontal dependence. The identity of our spheroidal coordinates
with spherical coordinates near the center insures the regularity of the
radial functions at the origin\footnote{We recall that a scalar function
expanded over the spherical harmonics basis $f(r,\theta,\varphi) =
\sum_{\ell,m}f_\ell^m(r)\Yl$ has radial components that verify
$f_\ell^m(r)\sim r^\ell$ when $r\tv0$.}.
Hereafter, $N_r$ will refer to the number of points of the radial
Gauss-Lobatto grid for the whole star, L$_{\rm max}$ will refer to
the highest degree of spherical harmonics used in the solution and
$N_\theta$ to number of grid points in latitude. Because of the imposed
equatorial symmetry of the models, L$_{\rm max}\simeq2N_\theta$.

\begin{figure}
\centerline{
\includegraphics[width=0.49\linewidth,clip=true]{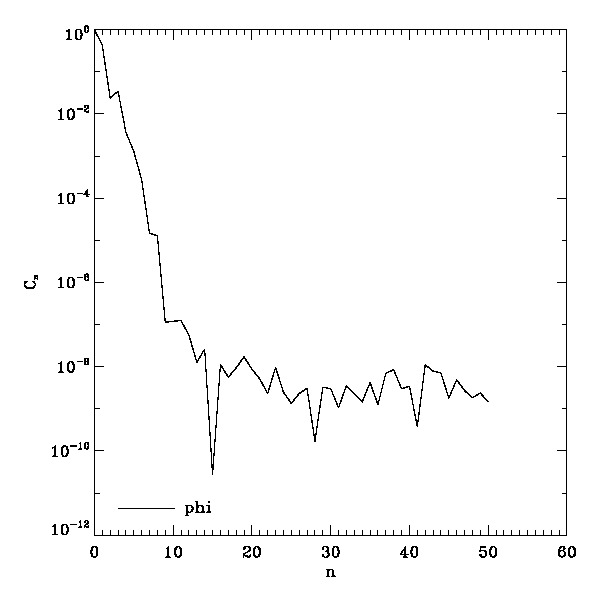}}
\caption[]{Spectrum of the gravitational potential from  the solution of
Poisson's equation for a n=3/2 polytrope. $C_n$ represents the absolute
value of the n$^{\rm th}$-Chebyshev coefficient. The star is covered by
a single domain.}
\label{spectpoly1D}
\end{figure}

The choice of a spectral representation is motivated by the precision
of spectral methods at a given resolution \cite[][]{CHQZ07}. Such
precision is required when eigenmodes and eigenfrequencies of a star
are computed and compared to observations. Previous computations of
polytropic stars (see below) with finite-difference schemes have shown
poor precision \cite[e.g.][]{AB94}.  For a virial test (see below) to
be met at a relative precision of $10^{-8}$, finite-differences need
5000 radial points \cite[][]{AB94}, while the same problem with the
same precision is solved with typically 20 points only when using the
Gauss-Lobatto collocation grid (see Fig.~\ref{spectpoly1D}). Our spectral
method with domain decomposition is based on the strong formulation of
the problem: equations or boundary/interface conditions are applied on
given collocation points without any overlapping of the domains unlike
the weak formulation \cite[e.g.][]{CHQZ07}. As previously mentioned,
the multidomain formulation turns out to be better adapted to the
strong variations of $\rho$ and provides more flexibility in the radial
distribution of grid points (see Fig.~\ref{spectra1D}).

Hence, the choice of a spectral element method turns out to be natural. We
note that except relativistic stars \cite[e.g.][]{BGM98}, stellar models
have never been computed with a spectral method. All one-dimensional
codes use finite-differences (like the MESA code of
\citealt{paxton_etal11} or the TGEC code, e.g. \citealt{hbh08}, etc.)
except the CESAM code, which is based on splines \cite[][]{Morel97,ML08}.

\section{The algorithm}

The existence of a converging algorithm for such a nonlinear and
complicated problem was the main uncertainty of the feasibility of this
modeling. We explored two algorithms, namely Picard and
Newton iteration schemes, which we shall now describe and discuss in the
context of our problem.

\subsection{Picard's iterative scheme}\label{picard}

Picard's algorithm is known as a very intuitive scheme for solving
iteratively nonlinear equation. It does only require the solving of a
linear equation that is already present in the physical problem at
hands. As such, it has been employed in many different problems:
flows in porous media \cite[][]{paniconiP94}, ice sheet flows
\cite[][]{lemieux_etal11}, MHD equilibria in tokamaks
\cite[][]{oliver_etal06}, and certainly many other problems. Although
we finally left this method for Newton's one, as actually done by all the
above mentioned works, it is worth mentioning the results we obtained
with it, at least to assess its efficiency and its limits in our stellar
modeling problem.

To present the results, we focus on the reduced problem of rigidly
rotating {\em polytropic
stars}. The general problem of modeling the steady state of rotating
stars, as represented by equations \eq{poisson}-\eq{angmom}, is strongly
simplified when one assumes a polytropic equation of state instead of
the general one\footnote{Polytropic stellar models are a valid option for
fully convective stars (i.e. stars with a mass less than 0.6 solar mass)
and for low-mass white dwarf stars for which the polytropic index
is 3/2. Due to their simplicity, these models have been the first to be
computed in two dimensions \cite[e.g.][]{james64}.}, namely if we assume
that the pressure only depends on the density through a power law like

\beq P = K\rho^{1+1/n}\eeqn{poly}
where $K$ is a constant and $n$ is the polytropic index.
In such a case the star may rotate as a solid-body\footnote{Non-uniform
rotation of polytropic stars have been explored by \cite{EM85},
\cite{AB94} and \cite{macgregor_etal07}.} and in the
appropriate rotating frame, one just needs to solve the hydrostatic
equation, namely

\beq -\na P -\rho\na\Phi_{\rm eff} = \vzero
\eeqn{hydrostat}
where $\Phi_{\rm eff}$ is the effective potential, namely the sum of
the gravitational potential $\Phi$ and the centrifugal potential $-\demi
\Omega^2 s^2$ ($\Omega$ is the angular velocity of the rigidly rotating
polytrope). From \eq{poly} and \eq{hydrostat}, it is easy to derive
the expression of density as a function of the gravitational
potential. Using scaled variables, Poisson's equation can be rewritten as

\beq \nabla^2\Phi = \lp 1 - \Lambda(\Phi-\Phi_0)
+\demi\Omega^2s^2\rp^n=\rho(\Phi) \eeqn{poissonpoly}
where $\Lambda=(1+\Omega^2/2)/(\Phi_{\rm
eq}-\Phi_c)$. In these expressions, $\Phi_{\rm eq}$ and $\Phi_c$ are the
equatorial and central values of the gravitational potential respectively.

For this ``simple" problem, we used a mapping with only two domains: one
for the star, the other for the surrounding vacuum (which we actually
limit to a sphere of radius $2R_{\rm eq}$). To use Picard's method
we also rewrite the Poisson equation $\nabla^2\Phi =\rho(\Phi)$ as

\beq\tilde{\nabla}^2\Phi = \frac{1}{g}\lp \rho(\Phi)+{\rm NS}\rp + \lp
1-\frac{g^{\zeta\zeta}}{g}\rp\tilde{\nabla}^2\Phi \eeqn{poissonPicard}
hence following \cite{BGM98}. In this expression $\tilde{\nabla}^2$
represents the spherical part of the Laplacian operator, namely

\[ \tilde{\nabla}^2 = \ddzeta{}+\frac{2}{\zeta}\dzeta{}
+\frac{\Delta_{\theta\varphi}}{\zeta^2} \with
\Delta_{\theta\varphi}=\dstheta{}+\ddsphi{},\]
and ``NS'' represents the non-spherical terms that, together with
$\tilde{\nabla}^2$, form the expression of the Laplacian in spheroidal geometry.
$g=\max(g^{\zeta\zeta})$ over each domain\footnote{The use of the global
maximum value is not recommended as we experienced a better convergence
when the maximum is evaluated on each subdomain.}. $g^{\zeta\zeta}$
is a component of the metric tensor (see appendix).

With Picard iterative scheme, equation \eq{poissonPicard} is solved in
the following way for the N+1$^{\rm th}$-iterate

\[ \tilde{\nabla}^2\Phi_{N+1} = \frac{1}{g_N}\lp{\rm NS}+\rho\rp_N +
\lp1-\frac{g^{\zeta\zeta}}{g}\rp_N\lp \lambda(\tilde{\nabla}^2\Phi)_N +
(1-\lambda)(\tilde{\nabla}^2\Phi)_{N-1}\rp\]
where $\lambda$ is a relaxation parameter usually set to 0.5 (but 1 is
found to work better here).

\begin{figure}
\centerline{
\includegraphics[width=0.49\linewidth,clip=true]{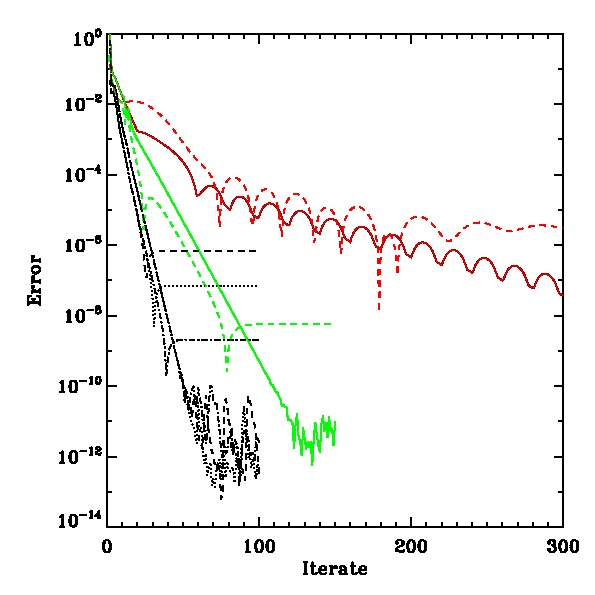}}
\caption[]{Convergence of the Picard iterations for the solution of the
rotating n=3/2-polytropic star. Black curves show the errors $\eps_1$
and $\eps_2$ of \eq{eps1} and \eq{eps2} as a function of the iteration
number for different spatial resolution in the $\Omega=0.5$ case: dashed
line $L_{\rm max}=8$ and Nr=32, dotted line $L_{\rm max}=16$ and Nr=32,
dashed-dotted line $L_{\rm max}=32$ and Nr=64. The curves terminating
by an horizontal plateau show the $\eps_2$-virial test. The two red curves
show the $\eps_1$ (solid) and $\eps_2$ (dashed) errors for a case
near criticality ($\Omega=0.77$ see table~\ref{table_perf}) and
with resolution $L_{\rm max}=64$ and Nr=128. The two green curves show
the same case as the red one, but when the grid is frozen at some stage
of the iterations.}
\label{converPicar}
\end{figure}

In Fig.~\ref{converPicar}, we show the convergence rate of the
iterations towards the solution for various spectral resolutions and for
two rotation rates, in the case of a polytropic index equal to 3/2.

To appreciate the convergence and the precision of the solution, we
examine two quantities: first the relative error on the gravitational
potential as a whole, namely

\beq \eps_1 = \frac{\max_{\rm star}(|\Phi_N-\Phi_{N-1}|)}{\max_{\rm
star}(|\Phi_N|)}\eeqn{eps1}
second, the error on the virial equality. Indeed, from the momentum
equation it can be shown that the exact solution must satisfy the
virial equality:

\[ I\Omega^2+W+3P = 0 \]
where $\Omega$ is the angular velocity of the polytrope, and

\[ I=\intvol r^2\sin^2\!\theta\, \rho dV, \qquad W = \demi\intvol\Phi\rho dV,
\qquad P = \intvol pdV\]
which represent physically the inertia moment, the gravitational binding
energy and the internal energy respectively. The quantity

\beq \eps_2= 1+\frac{I\Omega^2+3P}{W}\eeqn{eps2}
therefore measures the quality of the solution and incorporates the
three types of errors that plague the numerical solution: (i) the
truncation (spectral) error, iteration error and round-off errors.

\begin{figure}
\centerline{
\includegraphics[width=0.9\linewidth,clip=true]{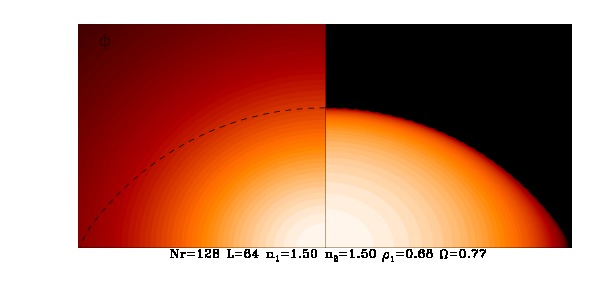}}
\caption[]{Distribution of the gravitational potential (left) and
density (right) for a n=3/2-polytropic star rotating near criticality
$\Omega/\Omega_{\rm crit}\simeq 0.98$. The ratio of the equatorial and polar
radii is $R_{\rm eq}/R_{\rm pol}=1.58$, corresponding to a flatness of
$\eps=0.367$.} 
\label{pot_rho}
\end{figure}

Let us now discuss the results. For the first rotation rate, $\Omega=0.5$
in our units, which corresponds to a configuration rotating at 68\% of
the critical angular velocity $\Omega_c=\sqrt{GM/R_{\rm eq}^3}$ and a
flatness\footnote{The flatness is defined as the ratio $(R_{\rm eq}-R_{\rm
pol})/R_{\rm eq}$. It measures the relative extension of the equatorial
radius compared to the polar one.} of 0.18, Fig.~\ref{converPicar}
shows that the convergence of the solution is rather fast. We observe
that whatever the spectral resolution, $\Phi$ converges exponentially
and reaches the round-off error after $\sim 50$ iterations. The level
of numerical noise is likely the results of the way the nonlinear terms
are computed and depends also on the condition number of the linear
operator. Quite remarkably, this effect depends very little on the
spatial resolution.

The virial error, is also sensitive to the spectral trunction. Thus it
first decreases as the error on the fields and then converges to the
spectral error. When spectral resolution is increased, this error is
decreased as expected.

The foregoing rotation rate ($\Omega=0.5$) represents rather fast rotating
stars, but some stars rotate near criticality. To test the efficiency
of the method we therefore set the rotation rate to $\Omega=0.77$,
which leads to a configuration that rotates at 98\% of the critical
angular velocity. As shown by Fig.~\ref{pot_rho}, at this angular
velocity the radius of curvature of the surface at equator is quite
small. Actually, at critical angular velocity, the star develops an
equatorial cusp, namely a non-smooth, discontinuous, variation of the
North-South tangent vector. This implies that stellar models rotating
near criticality are very demanding in angular resolution, and thus in
the spherical harmonic expansion. As shown in Fig.~\ref{converPicar}
(red pluses), convergence is now very slow. Actually, the algorithm
does not converge if a high precision on the solution is required. A
high precision solution at rotation rates close to criticality can
nevertheless be achieved (see table~\ref{table_perf}), by decoupling
the iterations on the grid from those on matter distribution. This is
indeed possible in the case of a polytropic star, since we are solving
for the gravitational potential, whose boundary conditions are set at
the star center and at infinity. Hence, even if the interface between
the domains does not match perfectly the true surface of the star,
this has little influence on the precision of the solution. Thus, when
computing the $\Omega/\Omega_c=0.99$ solution of table~\ref{table_perf},
we freezed the grid evolution when its flatness exceeds 30\%, otherwise
matter distribution slighly oscillates with the grid\footnote{Other
alternatives, like the use of a relaxation parameter, are possible,
but our trick turned out to be more expeditious.}. We note that a
similar trick has been used by \cite{gourgoulhon_etal01} for modeling
binary stars. In table~\ref{table_perf3}, we show similar results for
the n=3-polytrope.  We note that a much lower radial resolution is needed
to reach the required precision than with the n=3/2-polytrope. The reason
for that comes from the differentiability of the density near the surface
since $\rho \sim (R(\theta)-r)^n$ there. The high $n$ polytropes are
more easily represented with Chebyshev polynomials than the low $n$'s
\cite[see][for a thorough discussion of this point]{BGM98}.

\begin{table}[t]
\begin{center}
\begin{tabular}{ccccccccc}
\hline
         &                   &        &    &
&\multicolumn{2}{c}{Picard} & \multicolumn{2}{c}{Newton} \\
$\Omega$ & $\Omega/\Omega_c$ & $\eps$ & $L$& $N_r$ & N$_{\rm iter}$ &
CPU (s)&N$_{\rm iter}$& CPU (s)\\
         &                   &        &    &       &               &
&&\\
0.3      & 0.38              & 0.073  & 8  & 80+25 & 36            & 2
&15&2\\
0.5      & 0.64              & 0.183  & 14 & 90+25 & 40            & 3
&21&4.2\\
0.6      & 0.77              & 0.248  & 22 & 90+25 & 60            & 6
&23&7.9\\
0.7      & 0.90              & 0.317  & 42 & 90+25 & 137           & 23
&25&31\\
0.74     & 0.95              & 0.346  & 70 & 80+25 & 138           & 38
&27&92\\
0.76     & 0.98              & 0.361  & 120& 90+25 & 158           & 87
&27&600\\
0.772    & 0.99              & 0.370  & 200& 210+31& 120           & 373
&&\\
\hline
\end{tabular}
\caption[]{Characterization of the efficiency of the Picard and Newton
iterations for the computation of a n=3/2-polytrope at increasing rotation
rates.  $\Omega/\Omega_c$ gives the (uniform) angular velocity in terms of
the critical angular velocity and $\eps$ is the flatness of the rigidly
rotating polytropic star. $L$, $N_r$ and  N$_{\rm iter}$ are the minimum
angular, radial resolutions and number of iteration to reach a virial test
less than $10^{-9}$. CPU gives the CPU time in seconds needed for the
execution of the gfortran-compiled code on an Intel Core i5-5200U CPU @
2.20GHz. The last row at extreme rotation needed a slight change in the
Picard algorithm (see text), hence the lower number of iterations. For
Newton iterations we first start with the computation of a 1D-model and
continue with the desired 2D-model. We only give the number of iterations
needed by the 2D solutions, since they are the slowest.}
\label{table_perf}
\end{center}
\end{table}

\begin{table*}[t]
\begin{center}
\begin{tabular}{ccccccccc}
\hline
         &                   &        &    &
&\multicolumn{2}{c}{Picard} & \multicolumn{2}{c}{Newton} \\
$\Omega$ & $\Omega/\Omega_c$ & $\eps$ & $L$& $N_r$ & N$_{\rm iter}$ &
CPU (s) &N$_{\rm iter}$ &CPU (s)\\
         &                   &        &    &       &               & &&\\
0.2      & 0.44              & 0.069  & 8  & 28+13 & 50            & 0.5
&9& 1\\
0.3      & 0.66              & 0.151  & 14 & 28+15 & 64            & 1
&13& 1.3\\
0.35     & 0.77              & 0.203  & 20 & 28+17 & 75            & 1.2
&14& 1.5\\
0.41     & 0.90              & 0.275  & 48 & 28+19 & 136           & 6
&16& 4.2\\
0.43     & 0.95              & 0.302  & 80 & 28+19 & 130           & 13
&17& 10.8\\
0.445    & 0.98              & 0.323  & 176& 32+25 & 120           & 78
&20& 145\\
0.45     & 0.99              & 0.330  & 200& 32+25 & 140           & 127
&20& 187\\
\hline
\end{tabular}
\caption[]{Same as in table~\ref{table_perf} but for a n=3-polytrope.}
\label{table_perf3}
\end{center}
\end{table*}

\subsection{Newton's iterative scheme}

\subsubsection{Implementation}

Newton's algorithm solves iteratively a set of nonlinear equation
$\vF(\vX)=\vzero$ by assuming that a trial solution is close to the actual
solution so that a first order expansion of $\vF(\vX)$ can be used. If
$\vX_N$ is the N$^{\rm th}$ iterate then the correction $\delta\vX$
towards the true solution is obtained through

\[ \vF(\vX_N+\delta\vX)=\vzero \impl [J]\delta\vX = -\vF(\vX_N)\]
where $[J]$ is the Jacobian matrix of the nonlinear operator.

\begin{figure}
\centerline{
\includegraphics[width=0.5\linewidth,clip=true]{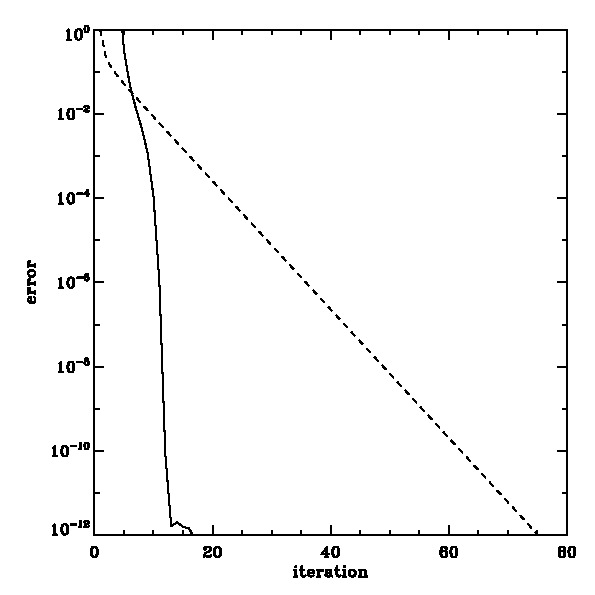}}
\caption[]{Convergence of Picard iterations (dashed line) and Newton
iterations (solid line) towards the solution of a non-rotating
polytropic star for n=3.5, using 100 radial grid points on a
Gauss-Lobatto grid.}
\label{newtpicard}
\end{figure}

We first tested Newton's method on the polytropic star without rotation,
thus in a one-dimensional situation to assess the good behaviour of the
scheme. We show the result in Fig.~\ref{newtpicard} along with the
result obtained with Picard iterations. Newton's method nicely
works and offers a fast (quadratic) convergence rate as expected.

The specificity of our problem is that the grid
is changing with the iterations and therefore the mapping
described in Sect.~\ref{sectmap} is also part of the variational
problem that defines the Jacobian. More precisely, the boundaries of the
domains are part of the vector $\vX$ defining the solution. Hence,
differential operators, which always use the metric tensor, should also
be differentiated with respect to  the grid variation. For instance, the
functional variation of the radial distance in the i$^{\rm th}$-domain
is given by

\[ \delta r^{i}= (\xi-A_i(\xi))\delta\Delta\eta_i + \delta
R_i(\theta) +  A_i(\xi)\delta\Delta R_i(\theta)\; ,\]
its $\zeta$-derivative by

\[ \delta r^{i}_\zeta = A'_i(\xi)\lp \frac{\delta
\Delta R_i(\theta)}{\Delta\eta_i} - \frac{\Delta
R_i(\theta)}{\Delta\eta_i^2}\delta\Delta\eta_i\rp\]
etc. These variations impact the variational form of the partial
differential equations. For instance, Poisson's equation yields

\[ \Delta\delta\phi + (\delta\Delta)\phi =
\delta\pi_c\rho+\pi_c\delta\rho\]
In this expression $\delta\Delta$ means the variation on the Laplacian
operator induced by the variation on the mapping. More explicitly,
from \eq{laplacien}, we find that

\[ \delta\Delta = \delta g^{\zeta\zeta}\ddzeta{}+\cdots \with \delta
g^{\zeta\zeta} = 2\frac{r\delta r + r_\theta\delta
r_\theta}{r^2r_\theta^2}-2g^{\zeta\zeta}\frac{r_\zeta\delta r +
r\delta r_\zeta}{rr_\zeta}\]
Hence, the expression of the Jacobian matrix is particularly cumbersome.

%In order to facilitate the evaluation of the elements of this matrix, we
%developed a symbolic parser that can automatically generate the terms of
%this matrix associated with a given term of the equations. For
%instance....

\bigskip
\begin{figure}
\centerline{
\includegraphics[width=0.5\linewidth,clip=true]{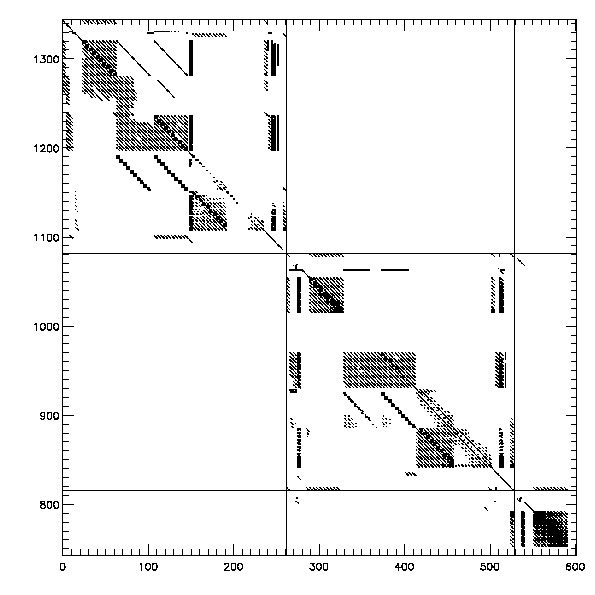}}
\caption[]{Shape of the Jacobian matrix (only first blocks shown). Grey
and black dots show non-zero elements while vertical and horizontal
lines delineate the position of the blocks. For this case $N_\theta=4$,
$N_r=10$ per domain.
}
\label{matrix}
\end{figure}

Once the Jacobian matrix is evaluated the linear system needs to be solved.
Typical resolutions need 300 grid points in the radial direction and 24
points in latitude (the star is assumed equatorially symmetric). Thus,
each physical scalar field generates a vector of about $10^4$ components
so that the needed five scalar fields together form a vector of few
$10^4$, thus a rather large, full matrix.

We solve this problem by a combination of a direct LU-solver
and the iterative Conjugate Gradient Squared (CGS) method
\cite[][]{sonneveld89}. Schematically, when the star is divided into
shell-like domains, the Jacobian matrix has a banded structure actually
made of coupled blocks of size determined by the resolution (see
Fig.~\ref{matrix}).
Since the CGS method needs preconditioning to be efficient, we first
LU-factorize the whole matrix taking advantage of its tri-diagonal
block structure. We thus generate a split pre-conditionner for the CGS
solver. However, the LU-factorization is expensive in terms of CPU time
so that the pre-conditionner is updated only when CGS iterations do not
converge within some fraction of the LU-factorization time. A similar kind
of algorithm has been used by \cite{EJ92} to solve 3D mixed convection
flows. Our algorithm may be summarized as follows:

\noindent
-----------------------------------------------------------------------
\begin{enumerate}

\item Read the initial model or build it (only in 1D)
\item Build the Jacobian matrix
\item LU-factorize the Jacobian matrix
\item CGS-solve for the correction $\delta\vX$
\item Update $\vX$
\item Recompute the RHS and the Jacobian matrix
\item Attempt another CGS solution to derive the new
$\delta\vX$ using the former LU matrices as a preconditionner. Namely,
if $L_k$ and $U_k$ are the factors of the Jacobian $\tens J_k$, the $k+1$
equation is solved with the (split-preconditioned) CGS solver as

\[ J_k^{k+1}U_k\delta\vX_{k+1} = -L_k^{-1}\vF_{k+1}  \]
where $J_k^{k+1}=L_k^{-1}\tens J_{k+1}U_k^{-1}$ is the preconditionned
matrix.  If the convergence of this iterative method is less than say N
iterates, then the algorithm continues at step 5. N is chosen such that
the N iterations are faster than a LU factorization. If convergence is
not reached, then the algorithm continues on step 3

\item if $\|\delta\vX\|/\|\vX\|\leq$tolerance, then stop.
\end{enumerate}
\noindent
-----------------------------------------------------------------------

\subsubsection{Test on the two-dimensional polytrope}

Since Picard method gives good results on the rotating 2D polytrope,
this problem offers a good set of comparisons for the methods.

When the iterations are started from scratch, Newton's scheme is much
slower than Picard's. The reason for that is the huge size of the
Jacobian matrix compared to the block-diagonal matrix of the Picard
iteration. Thus even if Newton's scheme uses much less iterations, their
cost immediately ruins its efficiency. Fortunately, Newton's
scheme can be easily improved if it first computes a 1D-model and use
this model as a start of the 2D-model. This much reduces the total CPU
times and Newton's method is then competitive.

The number of iterations and the needed CPU time for Newton's method are
given in tables~\ref{table_perf} and \ref{table_perf3}. From these
results we observe that Picard and Newton methods have a similar
efficiency when the rotation rate is less than the 90\% of the critical
one. Beyond this value, the required angular resolution is large and
Newton scheme has to deal with large full matrices which strongly slow
down the calculation. Hence, for the simple polytropic stars and for the
ultimate rotation rates, Picard's method appears to be the most
efficient\footnote{To be fair with Newton scheme, a further split of the
stellar domain into several subdomains may strongly accelerate the
iterations, at the price of some loss of precision of course.}.

\section{Numerical solution of stellar models}\label{stars}

We now turn to realistic stellar models as the one shown in
Fig.~\ref{rotdiff} or \ref{mercirc} and present their numerical
characteristics. Unless otherwise stated, we consider a 5 \msun\ stellar
model, with a chemically homogeneous composition close to the solar one
(X=0.7, Y=0.28 and Z=0.02). Such a model would describe a zero-age main
sequence B-star of our Galaxy.

\subsection{Picard's method}

The foregoing results let us think that Picard method should also work
with the complete set of equations (\ref{poissonnd} --
\ref{eq:angular_momnd}).
Unfortunately, this has not been the case. Using Kramers type opacities
(power laws) together with a simple equation of state (ideal gas), we
could not reach convergence when the angular velocity reached values
above $\Omega\sim0.35\Omega_c$. We ascribed this failure to the bad
conditioning of the linear operator, which now includes new coupling
terms, the worst of them coming from the momentum equation.

In view of the foregoing failure, we tried Newton's scheme on this
problem. We shall now present the results of this successful attempt.

\begin{figure}
\centerline{
\includegraphics[width=0.49\linewidth,clip=true]{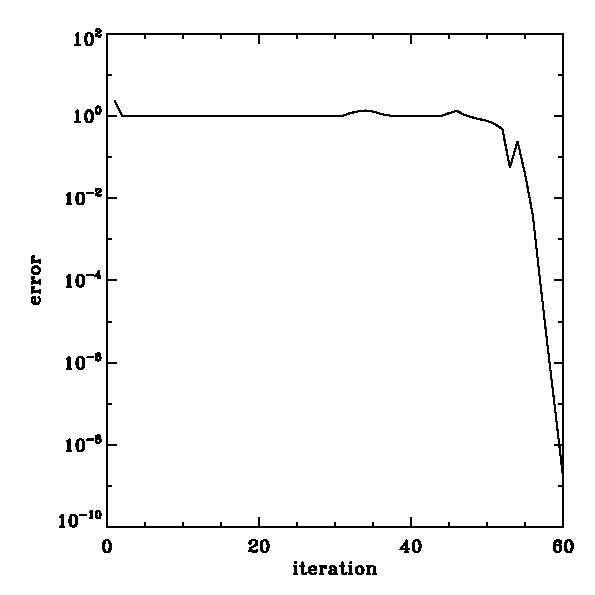}
\includegraphics[width=0.49\linewidth,clip=true]{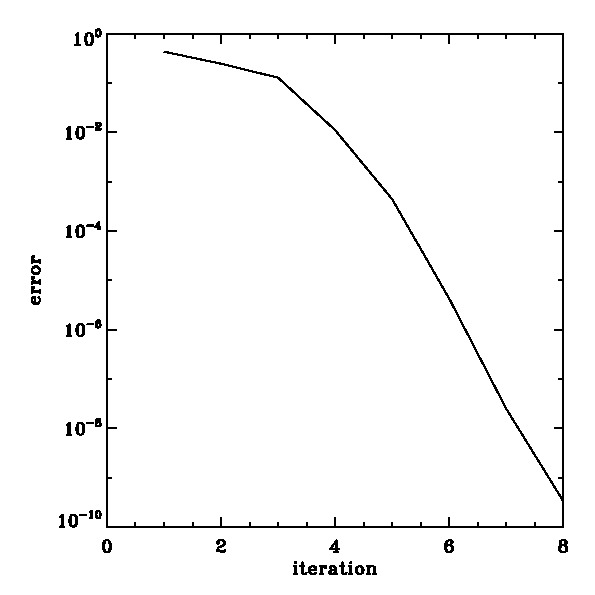}}
\caption[]{Left: Convergence of Newton's iterations for a 1D stellar model
started from scratch (5~\msun\ with homogeneous chemical composition -
see text). Right: Convergence of Newton's iterations for a 2D
model started from the converged 1D-model on left when
$\Omega/\Omega_c=0.66$.}
\label{convESTER12D}
\end{figure}

\begin{figure}
\centerline{
\includegraphics[width=0.85\linewidth,clip=true]{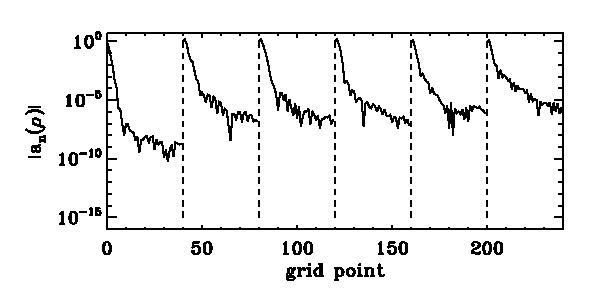}}
\centerline{
\includegraphics[width=0.85\linewidth,clip=true]{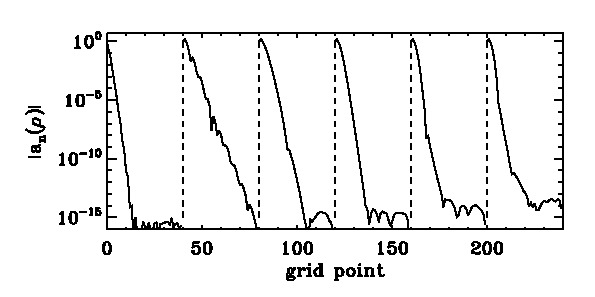}}
\centerline{
\includegraphics[width=0.85\linewidth,clip=true]{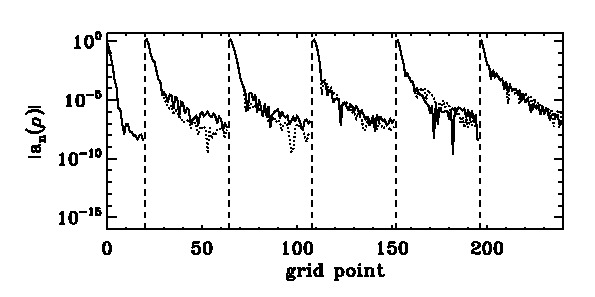}}
\caption[]{Top: Chebyshev spectrum in each domain for the density $\rho$
of a 1D-stellar model of 5\msun. We use 6 domains with 40 grid points
each.  Middle: similar model but where the equation of state is that of
the ideal gas and where the analytic Kramers type opacities expressions
are used. Bottom: Same as top but with a different distribution of grid
points: 20 points in the first domain (convective core) and 44 in the
remaining ones.  The spectral precision of the solution is typically
increased by an order of magnitude (see sect.\ref{stars}).  The
dotted lines show the spectra when Houdek \& Rogl's interpolations of
OPAL opacities are used \cite[][]{houdek_Rogl96}.}
\label{spectra1D}
\end{figure}

\subsection{One-dimensional models with Newton's scheme}

As is well-known, Newton iterations show quadratic convergence when the
initial guess is close enough to the actual solution. For our problem,
this implies that 2D models need to be initiated by a non-rotating
configuration of similar parameters just as for polytropes. We show in
Fig.~\ref{convESTER12D} the evolution of the relative L$_2$-norm
$\|\delta\vX\|/\|\vX\|$ of Newton's corrections with the iteration number
in the 1D and 2D cases. Since 1D iterations are extremely fast, they
are started from 'scratch' and are hence wandering around (typically
for 50 iterations) before the quadratic convergence operates. The
whole computation takes a few seconds on a laptop computer. The
result is then used to compute a 2D-model with $\Omega/\Omega_c=0.66$
(resulting in a flatness of 0.181), which shows a very good convergence
(e.g. Fig.~\ref{convESTER12D} right), showing that a converged 1D model
is an appropriate starting point for the computation of an already fast
rotating model.

Before discussing 2D-models, we shall first focus on the numerical
properties of the 1D models. Such models are indeed the first stellar models ever
produced using spectral methods. 

In Fig.~\ref{spectra1D} (top), we show the Chebyshev spectra, scaled by
the coefficient of the zeroth order polynomial, for each domain
used to compute the previous 5\msun\ 1D-model. The amplitude of the last
coefficient gives the relative spectral error (or truncation error) of
the Chebyshev polynomial representation. In such a realistic model, the
equation of state and the opacities are computed from interpolated
tables (OPAL). The spectral precision reflects the smoothness of
the interpolated functions and is clearly limited to $10^{-6}$. An
increased spectral resolution does not much improve the precision. Very
precise models are obtained when tabulated equation of state (EOS) and
tabulated opacities are replaced by -less realistic- analytical
formulae. For instance in Fig.~\ref{spectra1D} (middle) we use the ideal
gas EOS and Kramers opacity laws \cite[as in][]{ELR07}, namely

\beq P = \calR_* \rho T \andet \kappa = \kappa_0\rho^aT^b\eeqn{analy}
The solutions are then strikingly precise. We may also notice
that the central domain owns a better precision than the other
domains. The reason is that the central domain describes the convective
core of the star.  As discussed in section \ref{sectemp}, we assume  a constant
entropy in stellar cores.  Thus, numerical noise coming from opacity
disappears and we recover a better convergence similar to the case of
the polytrope (e.g. Fig.~\ref{spectpoly1D}). It is not as good as the
one obtained with \eq{analy} since the EOS remains more complex than
that of the ideal gas.

The better convergence in the central core allows us to use less radial
grid points there and to transfer them in the other domains. The precision
of the global solution can thus be improved as shown in
Fig.~\ref{spectra1D} (bottom). In this same figure we also tested
\cite{houdek_Rogl96} smooth interpolations of OPAL tables, but as shown,
the convergence is only slightly better in the central layers.

\begin{table}[t]
\begin{center}
\begin{tabular}{ccccccc}
\hline
$\Omega/\Omega_c$ & $\eps$ & $N_\theta$ & $N_r$ & N$_{\rm iter}$ & CPU
Time (s)& Precision\\
                  &        &    &           &            &        & \\
 0.70             & 0.200  & 16 & $6\times40$ &  8         & 170    &
$10^{-8}, 7\times10^{-4}$\\
 0.77             & 0.232  & 20 & $6\times40$ &  9         & 316 &
$2\times10^{-8}, 6\times10^{-4}$\\
 0.90             & 0.294  & 24 & $6\times40$ & +9         & +423 &
$4\times10^{-8}, 6\times10^{-4}$\\
 0.95             & 0.320  & 40 & $6\times40$ & +9         & +1200 &
$5\times10^{-8}, 3\times10^{-4}$\\
 0.98             & 0.337  & 40 & $6\times40$ & +11        & +1260 &
$4\times10^{-7}, 4\times10^{-4}$\\
 0.98             & 0.337  & 60 & $6\times40$ & +9         & +2150 &
$3\times10^{-8}, 5\times10^{-4}$\\
\hline
\hline
 0.77             & 0.233  & 20 & $10\times24$ & 10         & 203 &
$2\times10^{-8}, 9\times10^{-4}$\\
 0.90             & 0.295  & 24 & $10\times24$ & +9         & +217 &
$4.5\times10^{-8}, 7\times10^{-4}$\\
 0.90             & 0.295  & 24 & $10\times24$ & 13         & 456 &
$4.5\times10^{-8}, 7\times10^{-4}$\\
\hline
\end{tabular}
\caption[]{Some data showing the numerical efficiency of the ESTER
code at computing 5~\msun\ stellar models with fast rotation. $\Omega/\Omega_c$
is the fraction of critical angular velocity at equator. $\eps$ is the
flattening of the solution.  $N_\theta$ is the number of grid points
in latitude and $N_r$ the number of grid points inside the star in
the $\zeta$-coordinate. $N_r$ is given as the product of the number of
domains times the number of points inside a domain (here all domains
have the same resolution). CPU (s) gives the CPU-time needed for a
run with a i5-4570 processor at 3.2GHz.  Precision gives the numbers
corresponding to the virial and energy tests (see text).{\it Upper part:}
The $\Omega/\Omega_c=0.7$ and $\Omega/\Omega_c=0.77$ models are calculated
from a 1D model, while the following models (with $\Omega/\Omega_c\geq
0.90$) are computed from the preceding one. Hence, the 0.90-model is
iterated from the 0.77-model and needs 9-iterations using an additional
423s CPU time. The 0.95-model is computed from the 0.90-model with 9
additional iterations, etc. The 0.98-model was also computed with enhanced
resolution ($N_\theta=60$) from the 0.95-model. {\it Lower part:} Keeping
the same number of radial grid points, the number of domain is increased
to 10. A direct calculation of the 0.90-model is compared to a two-step
calculation using an intermediate 0.77-model.}
\label{table_perf2D}
\end{center}
\end{table}

\subsection{General properties of two-dimensional models}

Two-dimensional models are all computed from similar (in mass and chemical
composition) 1D-non-rotating models. We recall that 2D models need
very few input parameters, namely, the mass, the equatorial angular
velocity scaled by the keplerian one (i.e. $\Omega_{\rm eq}/\Omega_k$),
the hydrogen mass fraction in the envelope ($X$), in the core ($X_c$),
and the metallicity $Z$. Other parameters specify the numerical grid.
Thus, compared to 1D model, we just need to specify the equatorial rotation
rate or, equivalently, the total angular momentum.

From non-rotating models we can rather
easily compute a model rotating at 77\% of the critical angular velocity
as shown in Tab.~\ref{table_perf2D}. For a model, rotating at 90\% of
the critical velocity this is still possible but the number of domains
needs to be adjusted. Tab.~\ref{table_perf2D} shows two examples
of a computation of the 90\%-case: one directly from the non-rotating
model and the second with a first step at the 77\%-case. The two-step
calculation is slightly faster since the first steps can be done with
less angular resolution.

In Tab.~\ref{table_perf2D} we also give the accuracy of the solutions.
For that, we use the virial test as previously explained \cite[but see
also][]{REL13} and the energy test which demands that

\beq \intsur\chi\na T\cdot\dS + \intvol \eps_*dV = 0\; .\eeq
This latter test expresses the conservation of energy. The results given
in Tab.~\ref{table_perf2D} clearly show that this is satisfied to a
lower precision than the virial test. This is again an effect of the actual
numerical accuracy of the OPAL tables for opacity.

\begin{figure}
\centerline{
\includegraphics[width=0.70\linewidth,clip=true]{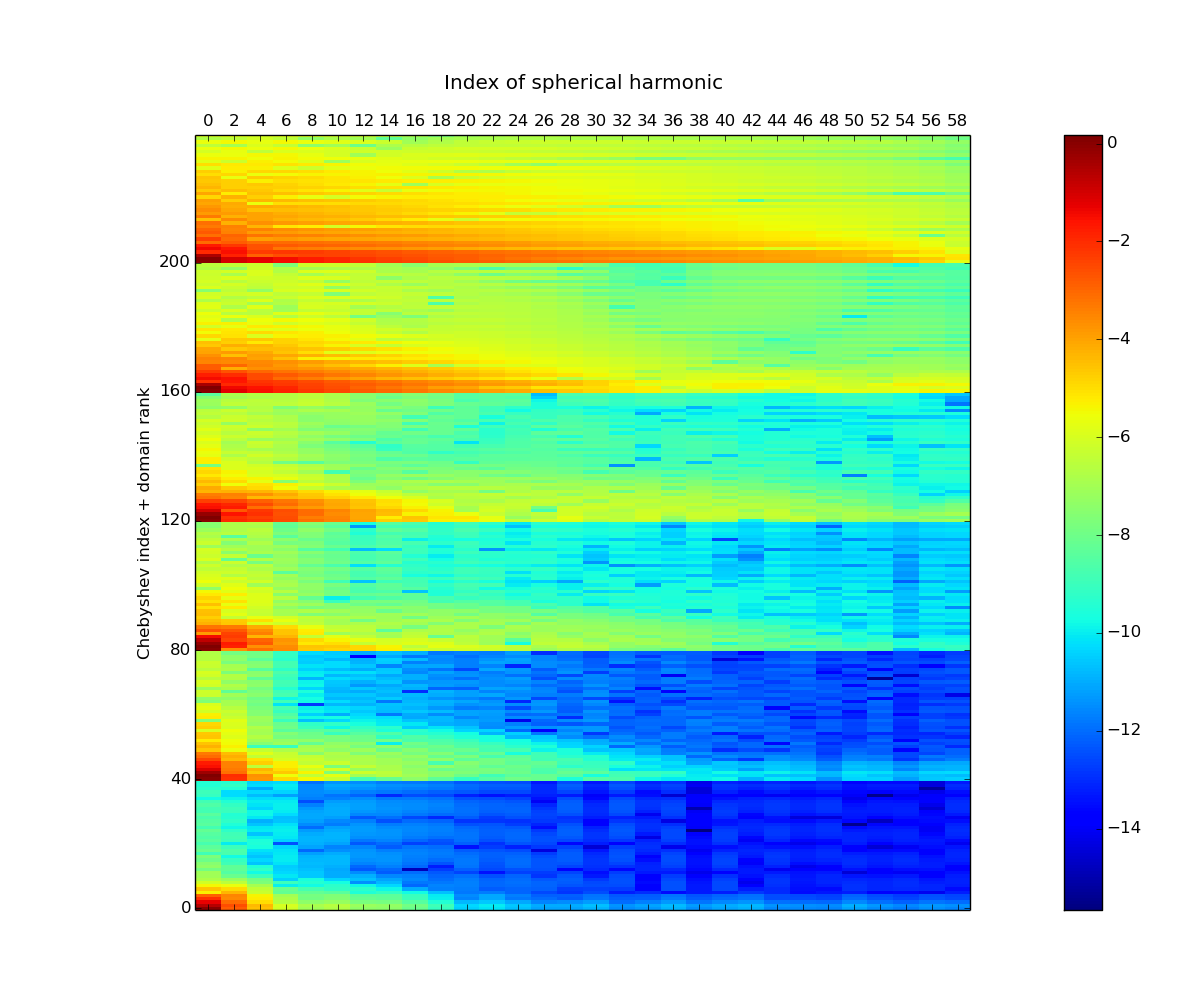}}
\caption[]{2D spectra of the density showing the convergence of the
solutions both in the spherical harmonic basis (horizontal axis)
and in the radial direction (vertical axis). In the radial direction,
each domain owns 40 grid points that give 40 Chebyshev coefficients
(the color scale is logarithmic). For this plot we used a 5~\msun\
model rotating at 95\% of the critical angular velocity.}
\label{spect2D}
\end{figure}

2D solutions must also satisfy spectral convergence in the horizontal
direction, namely on the spherical harmonic basis. In Fig.~\ref{spect2D},
we show a view of the corresponding 2D-spectrum.  Since the stellar domain
is divided into six ``radial" domains, we get six 2D-spectra. Putting them
side by side, we can appreciate the changes of the convergence with the
position of the domain in the star.  As in the 1D-case, surface layers are
the most difficult for spectral convergence. In these layers two effects
combine: the rapid variations of opacity and the more flattened shape
of outer domains compared with inner ones. This latter effect impacts
on the needed number of spherical harmonics.

\subsection{Numerical performances}

In the present version of the algorithm, the resolution of the linear system
associated with each Newton iteration is made through an LU factorization
and CGS iterations. The left and right preconditionning matrices needed by
the CGS algorithm
are obtained from the LU factorization of the Jacobian. This is an
efficient preconditioning since it allows convergence with 2 or 3 CGS
iterations. However, the LU factorization is an expensive operation whose
cost grows as $n^3$ compared to the matrix-vector products, or back
substitutions, that are used by the CGS scheme and which scale as $n^2$
($n$ is the size of the blocks). Thus, if $N_d$ is the number of domains,
$N_r$ the total number of radial grid points, and $N_\theta$ the number of
latitude grid points, the number of flops for each Newton iteration
increases as $N_r^3\times N_\theta^3/N_d^2$, since the LU factorization
dominates the costs. This expression readily shows that increasing the
number of domains is computationnaly advantageous. However, this is at the
expense of radial spectral precision (assuming that the total number of
radial grid points is fixed).

Actually, the optimal number of domains is such that radial and horizontal
spectral precision are about the same, which means that the number of grid
point per domain should be $\sim N^2_\theta$.  Table~\ref{table_perf2D}
shows that models calculations are affordable on a desktop/laptop computer
for models rotating up to 98\% of the critical angular velocity. Beyond
this value, the needed angular resolution gets very large because
the stellar surface tends to be singular (see Fig.~\ref{pot_rho} and
sect.~\ref{picard}). A cusp forms at the equator and other numerical
techniques need to be considered.

\section{Conclusions and outlooks}

In this paper, we presented the computational techniques of the ESTER
code, which computes the first stellar models based on spectral methods
and which can address the two-dimen\-sional case needed for fast rotating
stars.

The first key feature of the ESTER models is their use of a coordinate
system that is adapted to the shape of the star so that boundary
conditions can be easily imposed at the stellar surface. This is dealt
with a nonlinear mapping such that the new coordinate system reduces to
the spherical coordinates in the neighbourhood of the star's center. This
feature guarantees the regular behaviour of the radial spectral functions
at the origin.

A second key feature of these models is that they use shell domains with a
spectral decomposition inside each domain. Such a discretization of the
partial differential equations benefits from the high precision of spectral
methods and from the flexibility of domain distribution. This discretization
belongs to the class of spectral element methods \cite[][]{CHQZ07}.

The third key feature of the models is the choice of the algorithm for the
iterative method solving the nonlinear partial differential equations of
the problem. Both Picard and Newton schemes have been tested. The Picard
algorithm turns out to be very efficient on the simple polytropic models of
stars but loses its efficiency in stellar models with more complex physics.
On the contrary, Newton algorithm does not performs very well with simple
polytropes (in term of computing time), but is unsurpassed on realistic
stellar models with a complex microphysics (including tabulated data).
%These comparisons have been made without a refined tuning of the algorithm
%for the case at work. For instance, Newton algorithm efficiency could
%certainly be improved by fragmenting the stellar domain into multiple
%layers. However, the non-convergence of Picard scheme in the fast rotating
%case is setting out a much more difficult challenge.

The last key feature of our models is the combined use of LU factorization
and CGS iterations for the solving of Newton's iterations. The CGS
algorithm is crucial when 2D solutions are computed. It makes these
solutions affordable on small computers and does not jeopardize future
improvements including more complex physics.

This last point brings us to the possible and necessary future developments
of the ESTER code. These will have to deal with time evolution. But stellar
evolution is a long timescale process and therefore short timescale motion
like turbulence, should still be replaced by a mean-field modeling. Present 1D
models solve stellar evolution by chaining hydrostatic models in a
Lagrangian framework. Thus, modeling stellar evolution requires the
computation of thousands of models. In the two-dimensional framework, the
flows and the changing shape of the star calls for a mixt of the
Lagrangian and Eulerian formalism. The numerous models that will have to be
computed also demand that the numerical schemes are optimal. In this
respect, improving the parallelism of the code will be crucial. The
incomplete LU factorization is an interesting option for replacing the
present LU factorization which has not a great scalability. A more
challenging evolution of the code may be the use of Jacobian-free
Newton-Krylov methods which avoids the arduous work of deriving the
equations for the Newton corrections. However, for these methods,
preconditioning is also a strong issue \cite[][]{KK04}, and numerical
efficiency is not guaranteed. Thus, time-evolution will also require some
exploration to delineate the most appropriate numerical scheme, which is
presently unknown.

\bigskip
\noindent {\bf Acknowledgements}
\bigskip

We are very grateful to our colleague Daniel Reese for letting us
use his (very) illustrative figure (\ref{grid}), and to the referee
for very detailed and constructive remarks.  The authors acknowledge
the support of the French Agence Nationale de la Recherche (ANR),
under grant ESTER (ANR-09-BLAN-0140).  This work was also financially
supported by the Centre National de la Recherche Scientifique through
the Programme National de Physique Stellaire (PNPS, CNRS/INSU). The
numerical calculations have been carried out using HPC resources from
CALMIP (Grant 2015-P0107).

\bigskip
\noindent {\bf References}

\bigskip
\bibliographystyle{elsarticle-harv}
\bibliography{../../biblio/bibnew}

\appendix
\section{Some useful relations for the spheroidal geometry}

\subsection{Natural basis}

We start by defining the natural basis for the spheroidal coordinates. We
have two sets of basis vectors:

\begin{itemize}
\item Covariant basis vectors: $\displaystyle\vect{E}_i=\frac{\partial\vect r}
{\partial x^i}$
\begin{equation}
\vect E_\zeta=r_\zeta\rvec, \quad
\vect E_\theta=r_\theta\rvec+r\thvec, \quad
\vect E_\varphi=r\sin\theta\phivec,
\end{equation}
\item Contravariant basis vectors: $\vect E^i=\nabla x^i$
\begin{equation}
\vect E^\zeta=\frac{\rvec}{r_\zeta}-\frac{r_\theta}{rr_\zeta}\thvec, \quad
\vect E^\theta=\frac{\thvec}{r},\quad
\vect E^{\varphi}=\frac{\phivec}{r\sin\theta},
\end{equation}
\end{itemize}
where $\rvec,\thvec,\phivec$ are the usual unit vectors in
spherical coordinates, and $$r_\zeta=\frac{\partial r}{\partial\zeta}\quad
r_\theta=\frac{\partial r}{\partial\theta}$$

\begin{figure}[t]
\centerline{\includegraphics[width=5cm]{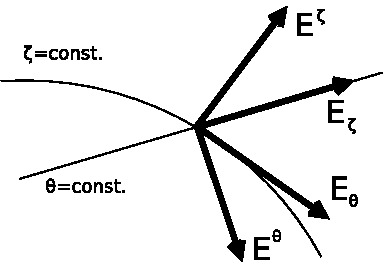}}
\caption[]{Contravariant and covariant vectors of the meridional plane
for the stellar models.}
\label{vectors}
\end{figure}

The vectors of the natural basis are not unit vectors. The covariant
vector $\vect E_i$ is parallel to the line $x^j=\cst$, with
$j\ne i$, while the contravariant vector $\vect E^i$ is perpendicular
to the surface $x^i=\cst$. For orthogonal coordinates $\vect
E_i$ and $\vect E^i$ are parallel, but this is not the case for non-orthogonal
coordinates. In fig.~\ref{vectors}, we sketch out the $\vect E_i$ and
$\vect E^i$ vectors in a meridional plane of the star.

The basis vectors satisfy
\begin{equation}
\vect E_i\cdot\vect E^j=\vect E^i\cdot\vect E_j=\delta_{ij}
\end{equation}
where $\delta_{ij}$ is the Kronecker's delta.

Using the basis vectors, we can calculate the metric tensor
\begin{equation}
g_{ij}=\vect E_i\cdot\vect E_j=\left(
\begin{array}{ccc}
r_\zeta^2&r_\zeta r_\theta&0\\
r_\zeta r_\theta&r^2+r_\theta^2&0\\
0&0&r^2\sin^2\theta
\end{array}
\right)
\end{equation}
or, in contravariant form
\begin{equation}
g^{ij}=\vect E^i\cdot\vect E^j=\left(
\begin{array}{ccc}
\displaystyle\frac{r^2+r_\theta^2}{r^2r_\zeta^2}&\displaystyle\frac{-r_\theta}{r^2r_\zeta}&0\\
\displaystyle\frac{-r_\theta}{r^2r_\zeta}&\displaystyle\frac{1}{r^2}&0\\
0&0&\displaystyle\frac{1}{r^2\sin^2\theta}
\end{array}
\right)
\end{equation}
Note that $g^{ij}$ is the matrix inverse of $g_{ij}$, namely
$$g_{ij}g^{jk}=\delta_{ij}$$
where we have used the Einstein's summation convention, that implies summation over repeated indices.

Given two points $x^i$ and $x^i+\mathrm{d}x^i$, the distance ($\mathrm{d}s$) between them 
is given by the metric tensor:
\begin{equation}
\mathrm{d}s^2=g_{ij}\mathrm{d}x^i\mathrm{d}x^j=
r_\zeta^2\mathrm{d}\zeta^2+2r_\zeta r_\theta \mathrm{d}\zeta \mathrm{d}\theta+
(r^2+r_\theta^2)\mathrm{d}\theta^2+r^2\sin^2\theta \mathrm{d}\varphi^2
\end{equation}

The basis vectors verify
\begin{equation}
\vect E_i\cdot(\vect E_j\times\vect E_k)=\epsilon_{ijk}
\end{equation}
and
\begin{equation}
\vect E^i\cdot(\vect E^j\times\vect E^k)=\epsilon^{ijk}
\end{equation}
where $\epsilon^{ijk}$ is the Levi-Civita tensor
\begin{equation}
\epsilon_{ijk}=\sqrt{|g|}[i,j,k]
\end{equation}
\begin{equation}
\epsilon^{ijk}=\frac{1}{\sqrt{|g|}}[i,j,k]
\end{equation}
where $|g|=\det(g_{ij})=r^4r_\zeta^2\sin^2\theta$ and
\begin{equation}
[i,j,k]=\left\{
\begin{array}{ll}
1&\quad\mbox{the arguments are an even permutation of
$\zeta,\theta,\varphi$}\\
-1&\quad\mbox{the arguments are an odd permutation of
$\zeta,\theta,\varphi$}\\
0&\quad\mbox{two or more arguments are equal}
\end{array}
\right.
\end{equation}

\subsection{Representation of vectors}

A vector $\vect v$ can be represented either in covariant or contravariant form:
\begin{itemize}
\item Covariant form: $\vect v=V_\zeta\vect E^\zeta+V_\theta\vect E^\theta+V_\varphi\vect E^\varphi$
\item Contravariant form: $\vect v=V^\zeta\vect E_\zeta+V^\theta\vect E_\theta+V^\varphi\vect E_\varphi$
\end{itemize}
Here, $V_i$ are the covariant components of the vector $\vec v$ and $V^i$ the contravariant components. Note that
$$\vect E_i\cdot\vect v=V_i \qquad \mbox{and} \qquad \vect E^i\cdot\vect v=V^i$$
We can use the metric tensor to pass from one representation to the other, indeed
\begin{equation}
V_i=\vect E_i\cdot\vect v=\vect E_i\cdot(\vect E_j V^j)=g_{ij}V^j
\end{equation}
and similarly
\begin{equation}
V^i=g^{ij}V_j
\end{equation}

Let $(v_r,v_\theta,v_\varphi)$ be the spherical
components of a vector $\vect v$ such that $\vect
v=v_r\rvec+v_\theta\thvec+v_\varphi\phivec$. Its spheroidal components
will be

\begin{equation}
V_\zeta=r_\zeta v_r,\quad V_\theta=r_\theta v_r+r v_\theta,\quad
V_\varphi=r\sin\theta v_\varphi
\end{equation}
and
\begin{equation}
V^\zeta=\frac{v_r}{r_\zeta}-\frac{r_\theta}{rr_\zeta}v_\theta,\quad
V^\theta=\frac{v_\theta}{r},\quad V^\varphi=\frac{v_\varphi}{r\sin\theta}
\end{equation}
We can see from this expressions that $V^\theta$ an $V^\varphi$ are in fact angular velocities.

Using the properties of the basis vectors it can be shown that the scalar product of two vectors is
given by
\begin{equation}
\vect a\cdot\vect b=A_iB^i=A^iB_i
\end{equation}
and the cross product is
\begin{equation}
\begin{array}{l}
(\vect a\times\vect b)^i=
\epsilon^{ijk}A_jB_k\\
(\vect a\times\vect b)_i=
\epsilon_{ijk}A^jB^k
\end{array}
\end{equation}

After the presentation of the basics of the representation of vectors in
spheroidal coordinates, let's see now a little example. Consider a
surface $\mathcal S$ defined by $\zeta=\cst$ as for example
the surface of a star or the frontier between two subdomains. We want to
calculate the normal and tangential projections of a vector $\vect v$ with
respect to $\mathcal{S}$. First, we define a unit vector $\vect{\hat n}$,
perpendicular to $\mathcal{S}$. For that, we just recall that $\vect
E^\zeta$ is perpendicular to the surfaces $\zeta=\cst$,
but it is not a unit vector, so

\begin{equation}
\vect{\hat n}=\frac{\vect{E^\zeta}}{|\vect{E^\zeta}|}=
\frac{\vect{E^\zeta}}{\sqrt{\vect{E^\zeta}\cdot\vect{E^\zeta}}}=
\frac{\vect{E^\zeta}}{\sqrt{g^{\zeta\zeta}}}
\end{equation}
then, the normal projection is
\begin{equation}
\vect{\hat n}\cdot\vect{v}=\frac{V^\zeta}{\sqrt{g^{\zeta\zeta}}}=
\frac{r_\zeta V^\zeta}{\sqrt{1+\frac{r_\theta^2}{r^2}}}
\end{equation}
For the parallel projection we have two vectors, the first one, in the
direction of $\varphi$ is just the spherical unit vector $\phivec$,
in the latitudinal direction, however, it will be

\begin{equation}
\vect{\hat t}=\frac{\vect{E_\theta}}{|\vect{E_\theta}|}=
\frac{\vect{E_\theta}}{\sqrt{\vect{E_\theta}\cdot\vect{E_\theta}}}=
\frac{\vect{E_\theta}}{\sqrt{g_{\theta\theta}}}
\end{equation}
so, the parallel projections over $\mathcal{S}$ are

\begin{equation}
\vect{\hat t}\cdot\vect{v}=\frac{V_\theta}{\sqrt{g_{\theta\theta}}}=
\frac{1}{\sqrt{1+\frac{r_\theta^2}{r^2}}}\frac{V_\theta}{r}
\end{equation}
and
\begin{equation}
\phivec\cdot\vect{v}=\frac{V_\varphi}{r\sin\theta}
\end{equation}

\subsection{Tensors}

A second order tensor $\tens T$ is represented using 2 indices
\begin{equation}
\tens T=T^{ij}\vect E_i\vect E_j=T_{ij}\vect E^i\vect E^j={T^i}j\vect E_i\vect E^j
={T_i}^j\vect E^i\vect E_j
\end{equation}
Again, we can use the metric tensor to lower and raise indices
\begin{equation}
\begin{array}{l}
T^{ij}=g^{ik}{T_k}^j=g^{jl}{T^i}_l=g^{ik}g^{jl}T_{kl}\\
T_{ij}=g_{ik}{T^k}_j=g_{jl}{T_i}^l=g_{ik}g_{jl}T^{kl}
\end{array}
\end{equation}
The tensor product of 2 vectors is a tensor

\begin{equation}
(\vect a\;\vect b)^{ij}=a^ib^j
\end{equation}
The dot product between a tensor and a vector is
\begin{equation}
(\tens T\cdot\vect v)^i=T^{ij}V_j
\end{equation}
and between a vector and a tensor
\begin{equation}
(\vect v\cdot\tens T)^j=T^{ij}V_i
\end{equation}
Finally, the double dot product is a scalar
\begin{equation}
\tens T : \tens T=T^{ij}T_{ij}
\end{equation}
All of this can be generalized to higher order tensors.

\subsection{Differential operators}

Our goal is to be able to write differential equations using spheroidal
coordinates. For that, we start finding the relation between the partial
derivatives with respect to the spherical coordinates and those calculated
with respect to the spheroidal coordinates. To clarify the notation,
we add a prime ($'$) to the spheroidal $\theta'$ and
$\varphi'$ coordinates. Thus, the derivative with respect to a spheroidal
coordinate is done holding the other spheroidal coordinates constant.
Following the chain rule

\begin{equation}
\frac{\partial}{\partial r}=
\frac{\partial\zeta}{\partial r}\frac{\partial}{\partial \zeta}
+\frac{\partial\theta'}{\partial r}\frac{\partial}{\partial \theta'}
+\frac{\partial\varphi'}{\partial r}\frac{\partial}{\partial \varphi'}
\end{equation}
Obviously, $\displaystyle \frac{\partial\theta'}{\partial
r}=\frac{\partial\varphi'}{\partial r}=0$, and

\[\mathrm{d}r=r_\zeta\mathrm{d}\zeta+r_\theta\mathrm{d}\theta, \qquad
\mathrm{d}\zeta=\frac{1}{r_\zeta}\mathrm{d}r-\frac{r_\theta}{r_\zeta}\mathrm{d}\theta\]
where we see that $\displaystyle \frac{\partial\zeta}{\partial
r}=\frac{1}{r_\zeta}$ and $\displaystyle
\frac{\partial\zeta}{\partial\theta}=-\frac{r_\theta}{r_\zeta}$. Then

\begin{equation}
\frac{\partial}{\partial r}=\frac{1}{r_\zeta}\frac{\partial}{\partial \zeta}
\end{equation}
The other partial derivatives are calculated in the same way

\begin{equation}
\frac{\partial}{\partial\theta}=\frac{\partial}{\partial \theta'}
-\frac{r_\theta}{r_\zeta}\frac{\partial}{\partial \zeta}
\andet \frac{\partial}{\partial \varphi}=\frac{\partial}{\partial \varphi'}
\end{equation}
Of course, we could substitute these expressions into the
expressions of the differential operators written with the spherical
coordinates, but there is a much more efficient way to do it.

First, let's define the general form of the gradient of a scalar
quantity. The gradient is a vector, whose covariant components are

\begin{equation}
(\nabla\phi)_i=\frac{\partial\phi}{\partial x^i}=\phi_{,i}
\end{equation}
where we have introduced the comma notation for the partial derivative. The contravariant components of 
the gradient will be
\begin{equation}
(\nabla\phi)^i=g^{ij}\phi_{,j}
\end{equation}

We can also derive a component of a vector $V^i$ in the same way. However, this derivative
\begin{equation}
\frac{\partial V^i}{\partial x^j}={V^i}_{,j}
\end{equation}
is not a tensor, as it does not transform as a tensor under a change
of coordinates. That's why one introduces the covariant derivative

\begin{equation}
\nabla_jV^i={V^i}_{;j}={V^i}_{,j}+{\Gamma^i}_{kj}V^k
\end{equation}
where $\displaystyle{\Gamma^i}_{kj}=\vect E^i\cdot\frac{\partial \vect
E_k}{\partial x^j}$ is a Christoffel symbol of the second kind. The
covariant derivative of a vector ${V^i}_{;j}$ is a tensor that represents
the gradient of the vector.

\begin{equation}
(\nabla\vect v)^{ij}=g^{jk}{(\nabla\vect v)^i}_k=g^{jk}{V^i}_{;k}
\end{equation}
We can also calculate the covariant derivative using the covariant
components of the vector

\begin{equation}
\nabla_jV_i={V_i}_{;j}={V_i}_{,j}-{\Gamma^k}_{ij}V_k
\end{equation}
The Christoffel symbols can be calculated using the following relation

\begin{equation}
\Gamma^i_{jk}=
\frac{1}{2}g^{il}(g_{lj,k}+g_{lk,j}-g_{jk,l})
\end{equation}
where we see that they are symmetric with respect to the second and
third indices $\Gamma^i_{jk}=\Gamma^i_{kj}$.  They also verify

\begin{equation}
\Gamma^{i}_{ji}=\frac{\partial\ln\sqrt{|g|}}{\partial x^j}
\end{equation}
The covariant derivative of second order tensors is obtained in a similar way

\begin{equation}
\nabla_kT^{ij}={T^{ij}}_{;k}={T^{ij}}_{,k}+{\Gamma^i}_{lk}T^{lj}+{\Gamma^j}_{lk}T^{il}
\end{equation}
If one of the indices is covariant, then we have

\begin{equation}
\nabla_k{T^i}_j={{T^i}_j}_{;k}={{T^i}_j}_{,k}+{\Gamma^i}_{lk}{T^l}_j-{\Gamma^l}_{jk}{T^i}_l
\end{equation}
where we can see the general rule valid also for higher order tensors, the covariant derivative is equal
to the regular derivative plus:
\begin{itemize}
\item for each contravariant index, $+{\Gamma^i}_{lk}T^{\ldots l\ldots}$
\item for each covariant index, $-{\Gamma^l}_{ik}T_{\ldots l\ldots}$
\end{itemize}

Using the covariant derivative, we can calculate all the differential
operators in spheroidal coordinates.  Hence,
the divergence of a vector is

\begin{equation}
\nabla\cdot\vect v=\nabla_iV^i={V^i}_{;i}
\end{equation}
and of a tensor

\begin{equation}
(\nabla\cdot\tens T)^i=\nabla_jT^{ij}={T^{ij}}_{;j}
\end{equation}
Note that some authors prefer the definition $(\nabla\cdot\tens
T)^j=\nabla_iT^{ij}={T^{ij}}_{;i}$, which is summed over the first index.
Using the expression for the cross product, we can calculate the curl of
a vector

\begin{equation}
(\nabla\times\vect v)^i=\epsilon^{ijk}\nabla_jV_k=\epsilon^{ijk}{V_k}_{;j}
\end{equation}
The Laplacian of a scalar field is

\begin{equation}
\nabla^2\phi=\nabla\cdot(\nabla\phi)=\nabla_i(g^{ij}\nabla_j\phi)=(g^{ij}\phi_{,j})_{;i}
\end{equation}
and for a vector field
\begin{equation}
(\nabla^2\vect v)^i=\nabla_j(g^{jk}\nabla_kV^i)=(g^{jk}{V^i}_{;k})_{;j}
\end{equation}
The material derivative is
\begin{equation}
\left[(\vect v\cdot\nabla)\vect v\right]^i=V^j\nabla_jV^i=V^j{V^i}_{;j}
\end{equation}

\subsection{Other relations}
\begin{itemize}
\item Line, area and volume elements
\begin{itemize}
\item Line element
\begin{equation}
\mathrm{d}s^2=g_{ij}\mathrm{d}x^i\mathrm{d}x^j=r_\zeta^2\mathrm{d}\zeta^2+2r_\zeta r_\theta \mathrm{d}\zeta \mathrm{d}\theta+
(r^2+r_\theta^2)\mathrm{d}\theta^2+r^2\sin^2\theta \mathrm{d}\varphi^2
\end{equation}
\begin{equation}
\mathrm{d}\vect r=\vect E_i \mathrm{d}x^i=\vect E_\zeta \mathrm{d}\zeta+\vect E_\theta \mathrm{d}\theta
+\vect E_\varphi \mathrm{d}\varphi
\end{equation}
\item Area element in a surface $\zeta=$const.
\begin{equation}
\mathrm{d}\vect S=(\vect E_\theta\times \vect E_\varphi)\mathrm{d}\theta \mathrm{d}\varphi=
r^2r_\zeta\sin\theta\vect E^\zeta \mathrm{d}\theta \mathrm{d}\varphi
\end{equation}
\begin{equation}
\mathrm{d}S=|\mathrm{d}\vect S|=
\sqrt{g^{\zeta\zeta}}r^2r_\zeta\sin\theta \mathrm{d}\theta \mathrm{d}\varphi=
r^2\sqrt{1+\frac{r_\theta^2}{r^2}}\sin\theta \mathrm{d}\theta \mathrm{d}\varphi
\end{equation}
\item Area element in a surface of constant $p=p(\zeta,\theta)$.
\begin{equation}
\mathrm{d}\vect S=
r^2r_\zeta\sin\theta\left(\vect E^\zeta+\frac{p_{,\theta}}{p_{,\zeta}}\vect E^{\theta}
\right) \mathrm{d}\theta \mathrm{d}\varphi
\end{equation}
\begin{equation}
\mathrm{d}S=|\mathrm{d}\vect S|=r^2r_\zeta\sin\theta
\sqrt{g^{\zeta\zeta}+2\frac{p_{,\theta}}{p_{,\zeta}}g^{\zeta\theta}
+\left(\frac{p_{,\theta}}{p_{,\zeta}}\right)^2g^{\theta\theta}}
\mathrm{d}\theta \mathrm{d}\varphi
\end{equation}
\item Volume element
\begin{equation}
\mathrm{d}V=\vect E_\zeta\cdot(\vect E_\theta\times\vect E_\varphi)\mathrm{d}\zeta \mathrm{d}\theta \mathrm{d}\varphi=
r^2r_\zeta\sin\theta \mathrm{d}\zeta \mathrm{d}\theta \mathrm{d}\varphi
\end{equation}
\end{itemize}
\item Differential operators
\begin{itemize}
\item Gradient
\begin{equation}
\nabla\phi=\phi_{,i}\vect E^i=\frac{\partial\phi}{\partial\zeta}\vect E^\zeta+
\frac{\partial\phi}{\partial\theta}\vect E^\theta+\frac{\partial\phi}{\partial\varphi}\vect E^\varphi
\end{equation}
\item Divergence
\begin{equation}
\begin{array}{rl}
\displaystyle\nabla\cdot\vect v&
\displaystyle ={V^i}_{;i}=
\frac{\partial V^i}{\partial x^i}+\frac{\partial\ln\sqrt{|g|}}{\partial x^k}V^k=\\
\\
&\displaystyle =\frac{\partial V^\zeta}{\partial\zeta}+\left(\frac{2r_\zeta}{r}
+\frac{r_{\zeta\zeta}}{r_\zeta}
\right)V^\zeta+\frac{\partial V^\theta}{\partial\theta}+\left(\frac{2r_\theta}{r}+
\frac{\cos\theta}{\sin\theta}+\frac{r_{\zeta\theta}}{r_\zeta}\right)V^\theta+
\frac{\partial V^\varphi}{\partial\varphi}
\end{array}
\end{equation}
\item Laplacian
\begin{equation}
\begin{array}{rl}
\nabla^2\phi&
\displaystyle =\Div(\nabla\phi)=(g^{ij}\phi_{,j})_{;i}=\frac{1}{\sqrt{|g|}}\frac{\partial}{\partial x^i}
\left(\sqrt{|g|}g^{ij}\frac{\partial\phi}{\partial x^j}\right)=\\
\\
&\displaystyle =
g^{\zeta\zeta}\frac{\partial^2\phi}{\partial\zeta^2}
+2g^{\zeta\theta}\frac{\partial^2\phi}{\partial\zeta\partial\theta}
+\frac{1}{r^2}\frac{\partial^2\phi}{\partial\theta^2}
+\frac{1}{r^2\sin^2\theta}\frac{\partial^2\phi}{\partial\varphi^2}
+\\
&\displaystyle\quad
+\left[\frac{2}{rr_\zeta}-\frac{r_{\theta\theta}}{r^2r_\zeta}
-g^{\zeta\zeta}\frac{r_{\zeta\zeta}}{r_\zeta}
-g^{\zeta\theta}\left(\frac{2r_{\zeta\theta}}{r_\zeta}-\frac{\cos\theta}{\sin\theta}\right)\right]
\frac{\partial\phi}{\partial\zeta}
+\frac{\cos\theta}{r^2\sin\theta}\frac{\partial\phi}{\partial\theta}
\end{array}
\label{laplacien}
\end{equation}
\item Curl
\begin{equation}
\begin{array}{rl}
\displaystyle \nabla\times\vect v &
\displaystyle = \epsilon^{ijk}V_{k;j}\vect E_i=\\
\\
&\displaystyle = \frac{1}{r^2r_\zeta\sin\theta}\left[
\left(\frac{\partial V_\varphi}{\partial\theta}-
\frac{\partial V_\theta}{\partial\varphi}\right)\vect E_\zeta +
\left(\frac{\partial V_\zeta}{\partial\varphi}-
\frac{\partial V_\varphi}{\partial\zeta}\right)\vect E_\theta +
\left(\frac{\partial V_\theta}{\partial\zeta}-
\frac{\partial V_\zeta}{\partial\theta}\right)\vect E_\varphi\right]\\
\end{array}
\end{equation}
\item Material derivative
\begin{equation}
\begin{array}{l}
\displaystyle(\vect a\cdot\nabla)\vect b
 =A^j{B^i}_{;j}\vect E_i=\\
\\
\quad\displaystyle =\left[A^\zeta\frac{\partial B^\zeta}{\partial\zeta}+
A^\theta\frac{\partial B^\zeta}{\partial\theta}+A^\varphi\frac{\partial B^\zeta}{\partial\varphi}
+\frac{r_{\zeta\zeta}}{r_\zeta}A^\zeta B^\zeta+\left(\frac{r_{\zeta\theta}}{r_\zeta}-\frac{r_\theta}{r}
\right)\left(A^\zeta B^\theta+A^\theta B^\zeta\right)+\right.\\
\quad\displaystyle \left. +\frac{1}{r_\zeta}\left(r_{\theta\theta}-
\frac{2r_\theta^2}{r}-r\right)A^\theta B^\theta+\frac{\sin\theta}{r_\zeta}\left(
r_\theta\cos\theta-r\sin\theta\right)A^\varphi B^{\varphi}\right]\vect E_\zeta +\\
\\
\quad\displaystyle +\left[A^\zeta\frac{\partial B^\theta}{\partial\zeta}+
A^\theta\frac{\partial B^\theta}{\partial\theta}+A^\varphi\frac{\partial B^\theta}{\partial\varphi}
+\frac{r_\zeta}{r}\left(A^\zeta B^\theta+A^\theta B^\zeta\right)
+\frac{2r_\theta}{r}A^\theta B^\theta-
\sin\theta\cos\theta A^\varphi B^{\varphi}\right]\vect E_\theta +\\
\\
\quad\displaystyle +\left[A^\zeta\frac{\partial B^\varphi}{\partial\zeta}+
A^\theta\frac{\partial B^\varphi}{\partial\theta}+A^\varphi\frac{\partial B^\varphi}{\partial\varphi}
+\frac{r_\zeta}{r}\left(A^\zeta B^\varphi+A^\varphi B^\zeta\right)
+\left(\frac{r_\theta}{r}+\frac{\cos\theta}{\sin\theta}\right)
\left(A^\theta B^\varphi+A^\varphi B^\theta\right)\right]\vect E_\varphi\\
\end{array}
\end{equation}
\end{itemize}

\item Christoffel symbols (different from 0)
\begin{equation}
\begin{array}{lll}
\displaystyle
\Gamma^{\zeta}_{\zeta\zeta}=\frac{r_{\zeta\zeta}}{r_\zeta}&
\displaystyle
\Gamma^{\zeta}_{\zeta\theta}=
\frac{r_{\zeta\theta}}{r_\zeta}-\frac{r_\theta}{r}&
\displaystyle
\Gamma^{\zeta}_{\theta\theta}=\frac{1}{r_\zeta}\left(r_{\theta\theta}-
\frac{2r_\theta^2}{r}-r\right)\\
\displaystyle
\Gamma^{\zeta}_{\varphi\varphi}=\frac{\sin\theta}{r_\zeta}(r_\theta\cos\theta
-r\sin\theta)&
\displaystyle
\Gamma^{\theta}_{\zeta\theta}=\frac{r_\zeta}{r}&
\displaystyle
\Gamma^{\theta}_{\theta\theta}=\frac{2r_{\theta}}{r}\\
\displaystyle
\Gamma^{\theta}_{\varphi\varphi}=-\sin\theta\cos\theta&
\displaystyle
\Gamma^{\varphi}_{\zeta\varphi}=\frac{r_\zeta}{r}&
\displaystyle
\Gamma^{\varphi}_{\theta\varphi}=\frac{r_\theta}{r}+
\frac{\cos\theta}{\sin\theta}
\end{array}
\end{equation}
where $\Gamma^i_{jk}=\Gamma^i_{kj}$ should be used for the remaining
ones.
\end{itemize}

\end{document}